\documentclass[10pt,twocolumn]{article}
\usepackage[top=1.9cm, bottom=1.9cm, left=1.8cm, right=1.8cm]{geometry}
\usepackage{times}  %
\usepackage[sort&compress,numbers]{natbib}  %
\usepackage[hyphens]{url}
\usepackage{graphicx}  %
\usepackage[keeplastbox]{flushend}
\usepackage{graphicx} %
\usepackage{tikzsymbols}

\let\oldbibliography\thebibliography
\renewcommand{\thebibliography}[1]{%
\oldbibliography{#1}%
\setlength{\itemsep}{2pt}%
}
\usepackage[hang,flushmargin]{footmisc}

\usepackage[compact]{titlesec}
\titlespacing*{\section}{0pt}{*3}{3pt}
\titlespacing*{\subsection}{0pt}{*2}{2pt}
\usepackage{xspace}
\makeatletter
\def\url@leostyle{%
  \@ifundefined{selectfont}{\def\UrlFont{}}%
  {\def\UrlFont{}}%
}
\makeatother
\usepackage[hyphenbreaks]{breakurl}
\usepackage[bookmarks=true, bookmarksnumbered=true, colorlinks=true, linkcolor=linkcol, citecolor=citecol, urlcolor=urlcol, hypertexnames=true]{hyperref}
\definecolor{darkgreen}{RGB}{0, 100, 0}
\definecolor{linkcol}{rgb}{0.3,0,0}
\definecolor{citecol}{rgb}{0.3,0,0}
\definecolor{urlcol}{rgb}{0.3,0,0}
\definecolor{vlightgray}{gray}{0.925}

\usepackage[small,bf]{caption}
\usepackage{tikz}
\usepackage{amsmath}
\usepackage[sort&compress]{natbib}
\usepackage{comment}
\usepackage{booktabs}
\usepackage{wrapfig}
\usepackage{xspace}
\usepackage{mdframed}
\usepackage{fdsymbol}

\newcommand{\knowntrolls}{335\xspace}
\newcommand{\expandeddataset}{53,763\xspace}
\newcommand{\detectedtrolls}{1,248\xspace}
\newcommand{\randomaccounts}{1,000\xspace}
\newcommand{\system}{\textsc{TrollMagnifier}\xspace}
\newcommand{\fscore}{97.8}

\newcommand{\rev}[1]{ \textcolor{black}{#1}}
\newcommand{\newrev}[1]{ \textcolor{black}{#1}}

\newif\ifcomment
\commentfalse
\ifcomment
\newcommand{\sz}[1]{{\bf \textcolor{brown}{SZ: #1}}}
\newcommand{\jbnote}[1]{{\bf \textcolor{magenta}{JB: #1}}}
\newcommand{\edc}[1]{{\bf \textcolor{blue}{EDC: #1}}}
\newcommand{\gs}[1]{{\bf \textcolor{red}{GS: #1}}}
\newcommand{\hs}[1]{{\bf \textcolor{orange}{HS: #1}}}
\else
\newcommand{\sz}[1]{}
\newcommand{\jbnote}[1]{}
\newcommand{\edc}[1]{}
\newcommand{\gs}[1]{}
\newcommand{\hs}[1]{}
\fi
\newcommand{\descr}[1]{\smallskip\noindent\textbf{#1}}

\usepackage{subfigure}
\makeatletter
\def\url@leostyle{%
  \@ifundefined{selectfont}{\def\UrlFont{\small}}%
  {\def\UrlFont{}}%
}
\makeatother
\begin{document}
\date{}

\title{\bf \system: Detecting State-Sponsored Troll Accounts\\on Reddit\footnotemark}

\author{Mohammad Hammas Saeed$^{\clubsuit}$, Shiza Ali$^{\clubsuit}$, Jeremy Blackburn$^{\vardiamondsuit}$, \\Emiliano De Cristofaro$^{\varheartsuit}$, Savvas Zannettou$^{\spadesuit \blacktriangle}$ and Gianluca Stringhini$^{\clubsuit}$\\[0.5ex]
\normalsize $^{\clubsuit}$Boston University, $^{\vardiamondsuit}$Binghamton University, $^{\varheartsuit}$University College London, \\
\normalsize$^{\spadesuit}$TU Delft, $^{\blacktriangle}$Max Planck Institute for Informatics\\
\normalsize \{hammas,shiza,gian\}@bu.edu, jblackbu@binghamton.edu, e.decristofaro@ucl.ac.uk, s.zannettou@tudelft.nl}

\maketitle

\begin{abstract}

Growing evidence points to recurring influence campaigns on social media, often sponsored by state actors aiming to manipulate public opinion on sensitive political topics.
Typically, campaigns are performed through instrumented accounts, known as {\em troll accounts}; %
despite their prominence, however, little work has been done to detect these accounts in the wild.
In this paper, we present \system, a detection system for troll accounts.
Our key observation, based on analysis of known Russian-sponsored troll accounts identified by Reddit, is that they show loose coordination, often interacting with each other to further specific narratives.
Therefore, troll accounts controlled by the same actor often show similarities that can be leveraged for detection.
\system learns the typical behavior of known troll accounts and identifies more that behave similarly.
We train \system on a set of \knowntrolls known troll accounts and run it on a large dataset of Reddit accounts.
Our system identifies \detectedtrolls potential troll accounts; we then provide a multi-faceted analysis to corroborate the correctness of our classification.
In particular, \rev{66\% of the detected accounts} show signs of being instrumented by malicious actors (e.g., they were created on the same exact day as a known troll, they have since been suspended by Reddit, etc.). 
They also discuss similar topics as the known troll accounts and exhibit temporal synchronization in their activity.
Overall, we show that using \system, one can grow the initial knowledge of potential trolls provided by Reddit by over 300\%.

\end{abstract}

\renewcommand{\thefootnote}{\fnsymbol{footnote}}
\footnotetext{$^*$A shorter version of this paper appears in the Proceedings of the 43th IEEE Symposium on Security and Privacy (Oakland'22). This is the full version.}

\section{Introduction}
	
Social media has dramatically changed the way in which people get and consume news~\cite{kwak2010twitter,lerman2010information}.
Alas, 
this has also facilitated the dissemination of misleading information (i.e., \emph{misinformation}) and of deliberate campaigns to spread false narratives (i.e., \emph{disinformation})~\cite{starbird2019disinformation,starbird2017examining,vosoughi2018spread,wang2020understanding}.
Disinformation campaigns are often orchestrated by state actors, with the goal of polarizing public discourse or pushing talking points to favor particular agendas~\cite{mueller2019mueller,nic2021foreign}.
To do so, malevolent actors instrument so-called \emph{troll accounts} to engage in discussions among each other and with real users, pushing certain narratives and sharing false information~\cite{xiadisinfo}.

Social network providers have been working to identify and suspend these accounts and released information about them after the fact~\cite{reddittrolls,twittertrolls}.
This has helped researchers shed light on how troll accounts were operated and the narratives they were pushing, in particular with respect to state-sponsored troll accounts active on Twitter and Reddit between 2014 and 2018~\cite{bessi2016social,ferrara2017disinformation,xiadisinfo,zannettou2019characterizing,zannettou2019disinformation,zannettou2019let}. %

\descr{Research Problem.} %
Detecting troll accounts is a more difficult task than %
detecting ``traditional'' automated malicious activity.
Unlike malicious accounts involved in fraud and spam, those taking part in influence campaigns are usually controlled, manually, by humans.
Miscreants craft a set of accounts and control them directly, posting messages, interacting with real users, and with each other.
This means that troll accounts do not show strict synchronization patterns that are typical of automated activity, and which were the foundation of previously proposed detection systems~\cite{cao2014uncovering,gao2010detecting,grier2010spam,nilizadeh2017poised,stringhini2015evilcohort}.
Additionally, they present traits that are more similar to regular users, %
thus making approaches that rely on identifying mass-created fake accounts or bulk content ineffective~\cite{benevenuto2010detecting,ferrara2016rise,stringhini2010detecting,yang2011free,gao2010detecting,grier2010spam,xudeep}.

Overall, while state-sponsored troll accounts exhibit some indicative traits, typically, social networks identify them via ad-hoc analysis.
Twitter and Reddit released information about thousands of troll accounts~\cite{reddittrolls,twittertrolls}, but they did not disclose the methods that they followed to identify them, and it is unclear how comprehensive these detections were.
In this paper, we aim to automatically detect state-sponsored troll accounts on Reddit.
Our intuition, informed by previous studies~\cite{starbird2019disinformation,volkova2016account,zannettou2019disinformation}, is that accounts controlled by the same entity work together to push certain disinformation narratives.
This loose coordination generates interaction patterns that are measurable, and can be used for detection.
For instance, troll accounts that belong to the same campaign might often reply to each other, or follow up to discussions started by other troll accounts to keep the discussion alive and attract real users.
Therefore, by learning these interaction patterns from a set of known troll accounts, it should be possible to identify more accounts controlled by the same state-sponsored actors. %

\descr{\system.} We present a system called \system;
we train it on a dataset of \knowntrolls Russian-sponsored troll accounts identified by Reddit, which were active on the platform between 2015 and 2018~\cite{reddittrolls}.
We first demonstrate that they do show peculiar interaction patterns compared to regular Reddit accounts. %
For example, troll accounts are more likely to reply to each other or to make submissions with the same title.
We identify several features that characterize troll accounts (e.g., the fraction of comments made on submissions by troll accounts or the fraction of submissions with the same title as a troll account's submission) 
and use them to train classifiers and identify additional troll accounts in the wild.

\descr{Results.} Our experimental analysis shows that \system can effectively distinguish troll and benign accounts on our labeled dataset with up to \fscore\% F1-Score.
We then run \system on unseen Reddit accounts: our system identifies \detectedtrolls as likely Russian-sponsored troll accounts.
To confirm our results, we perform additional analysis, showing that \rev{66\% of the detected accounts} were either later suspended by Reddit, deleted some of their comments of submissions (typical behavior of troll accounts observed by previous work~\cite{zannettou2019disinformation}), or were created on the same day of a known troll account.
We also find that the detected troll accounts show stronger timing coordination patterns in their activity to the set of known troll accounts, compared to undetected accounts, and that they tend to use similar language to the one used by the known troll accounts, indicating that they are likely controlled by the same actor.
Our results show that interaction patterns are an effective way to characterize and identify trolls on Reddit. %

\descr{Contributions.} This paper makes the following contributions:

\begin{itemize}
 \item We show that the interactions on Reddit of troll accounts from the same campaigns are quite different from those of regular accounts on the same subreddits, and that this can be used for detection.

 \item We develop \system, a system able to learn the typical behavior of a seed of known troll accounts and find more accounts that showed a similar behavior on Reddit.

 \item We run \system on Reddit accounts extracted from Pushshift~\cite{baumgartner2020pushshift}. 
 Out of \expandeddataset accounts that interacted with the known troll accounts, \system identifies \detectedtrolls potential troll accounts.
\item We perform a multi-faceted evaluation of our approach.
  We show that the accounts detected by \system present signs of being controlled by malicious actors.
We also perform qualitative experiments to determine \system's false negatives, estimating that the false negative rate of our approach is 10\%.

\end{itemize}

We shared our results with Reddit and are waiting for their feedback.

\section{Preliminaries}

In this section, we describe our threat model, then, we present the main characteristics of the Reddit social network.
Finally, we introduce the datasets used in this work.

\subsection{Threat Model}

Based on the observations made by previous analysis of state-sponsored troll accounts~\cite{ferrara2017disinformation,mueller2019mueller,xiadisinfo,zannettou2019disinformation,zannettou2019let}, we describe the operation of a typical troll campaign as follows:

\begin{enumerate}
  \item One or more malicious actors create and instrument a number of accounts -- which we denote as ``\emph{troll accounts}'' -- on a social network. 
  These are populated with data (e.g., profile pictures and profile description) that makes them look believable and fit the narrative that the malicious actor wants to push (e.g., a retired man from Southern England).
  \item The troll accounts act ``normally'' for a while, posting content not related to disinformation, with the goal of attracting followers.
 This is common for other types of malicious accounts as well, e.g., for spam~\cite{stringhini2010detecting}.
  \item The troll accounts start pushing specific narratives. %
    They post original content (e.g., links to news articles or posts containing false or manipulated pictures) and simulate discussion between each other.
    They also engage in conversations with legitimate users with the goal of derailing and polarizing the discussion~\cite{mueller2019mueller,zannettou2019let}.
  \item Unwitting legitimate accounts react to the content posted by the troll accounts, e.g., re-sharing it or interacting directly with them.
  This will turn the disinformation seeds planted by the malicious actor into an organic disinformation campaign where content is shared by both troll accounts and legitimate users~\cite{starbird2019disinformation}.
\item \gs{If we are desperate for space we can delete this}After the disinformation campaign is over, the malicious actor might reset the troll accounts, deleting their posts and/or changing their profile traits%
~\cite{zannettou2019disinformation}. %
    For example, Russian-sponsored troll accounts participated in disinformation campaigns about Crimea but later changed identity and started focusing on US-based political issues.
 This makes detection more difficult, because of the lack of visibility on past identities and deleted content.
\end{enumerate}

\subsection{Reddit}

Reddit is a popular news aggregation site, where content is organized into millions of user-generated communities, called \emph{subreddits}, covering topics of interest ranging from news and sports~\cite{weninger2013exploration} to conspiracy theories~\cite{samory2018conspiracies}.
A user can create a thread -- more specifically, a \emph{submission} -- and other users can reply in a structured manner by posting \emph{comments}.
That is, users can reply to the submission itself or to comments to the submission. %
We focus on Reddit because this platform has become popular among Internet users, in particular when discussing news~\cite{samory2018conspiracies,weninger2013exploration,weninger2014exploration,zannettou2017web}.
Also, the fact that posts on Reddit are organized in threads allows us to study the interaction between troll accounts, which is key to our approach.

\begin{figure*}[t]
\centering
\subfigure[Started]{\includegraphics[width=0.32\textwidth]{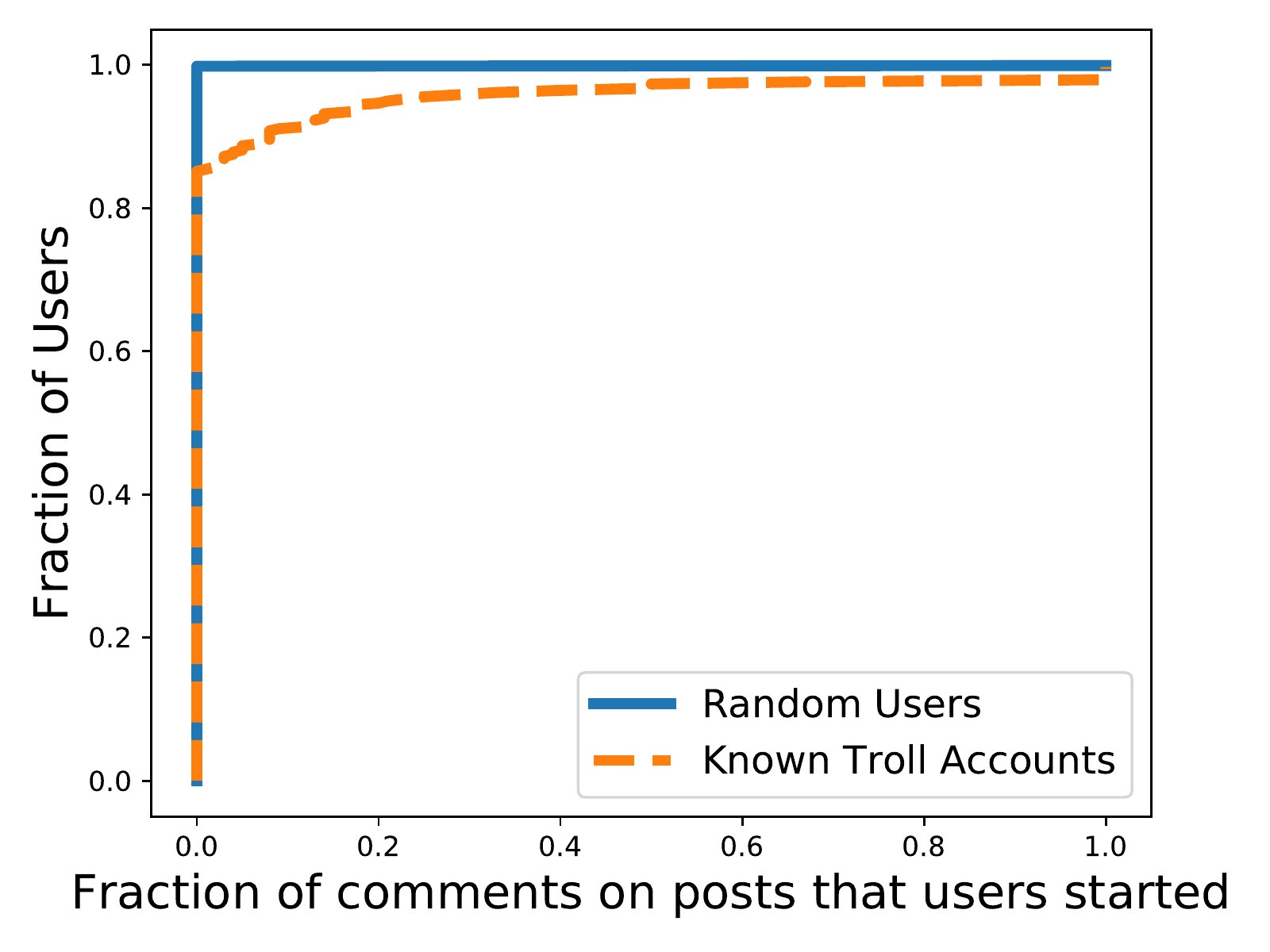}\label{fig:comments_started}}
\subfigure[Commented on]{\includegraphics[width=0.32\textwidth]{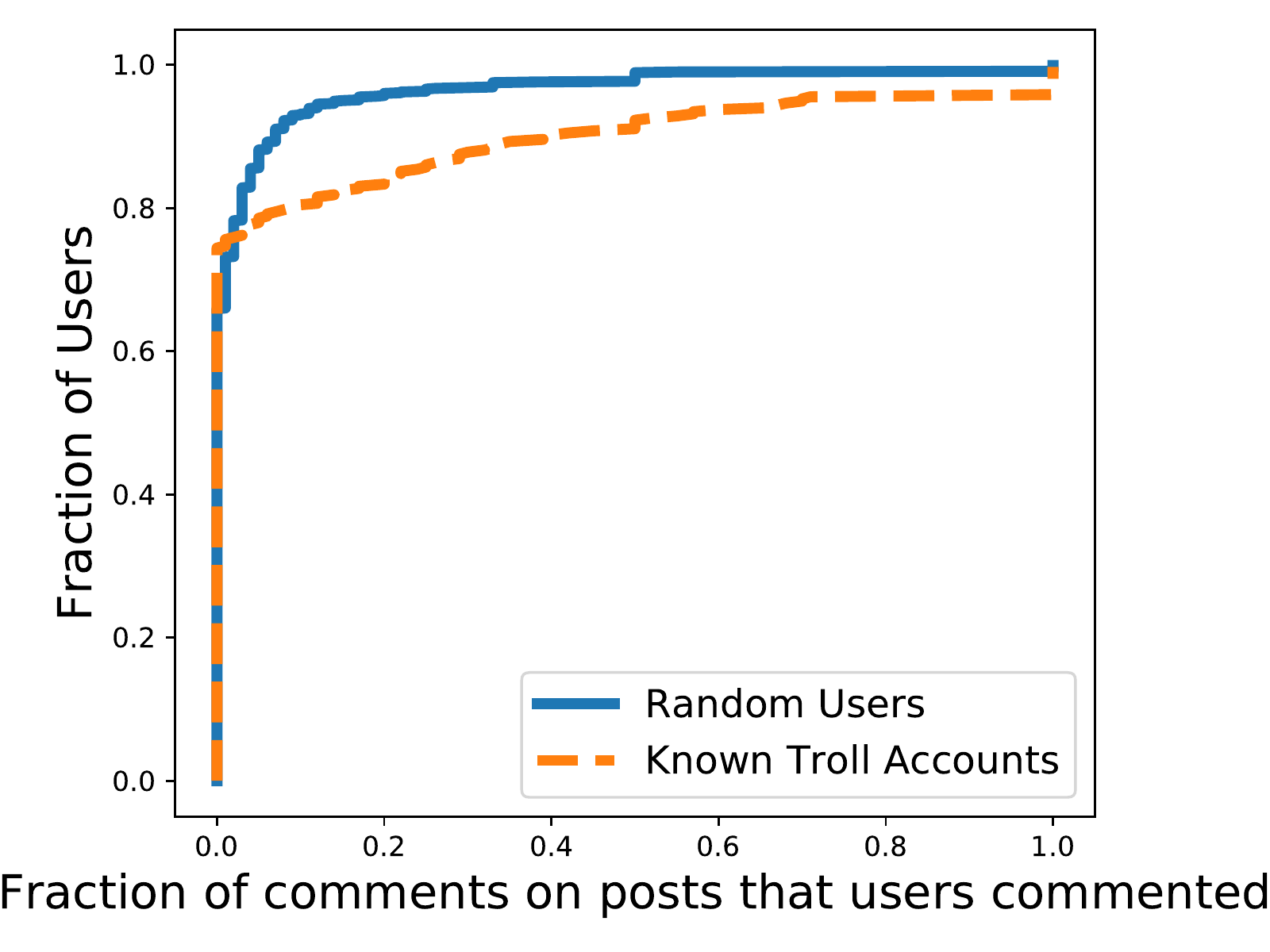}\label{fig:same_comments}}
\subfigure[Same Title]{\includegraphics[width=0.32\textwidth]{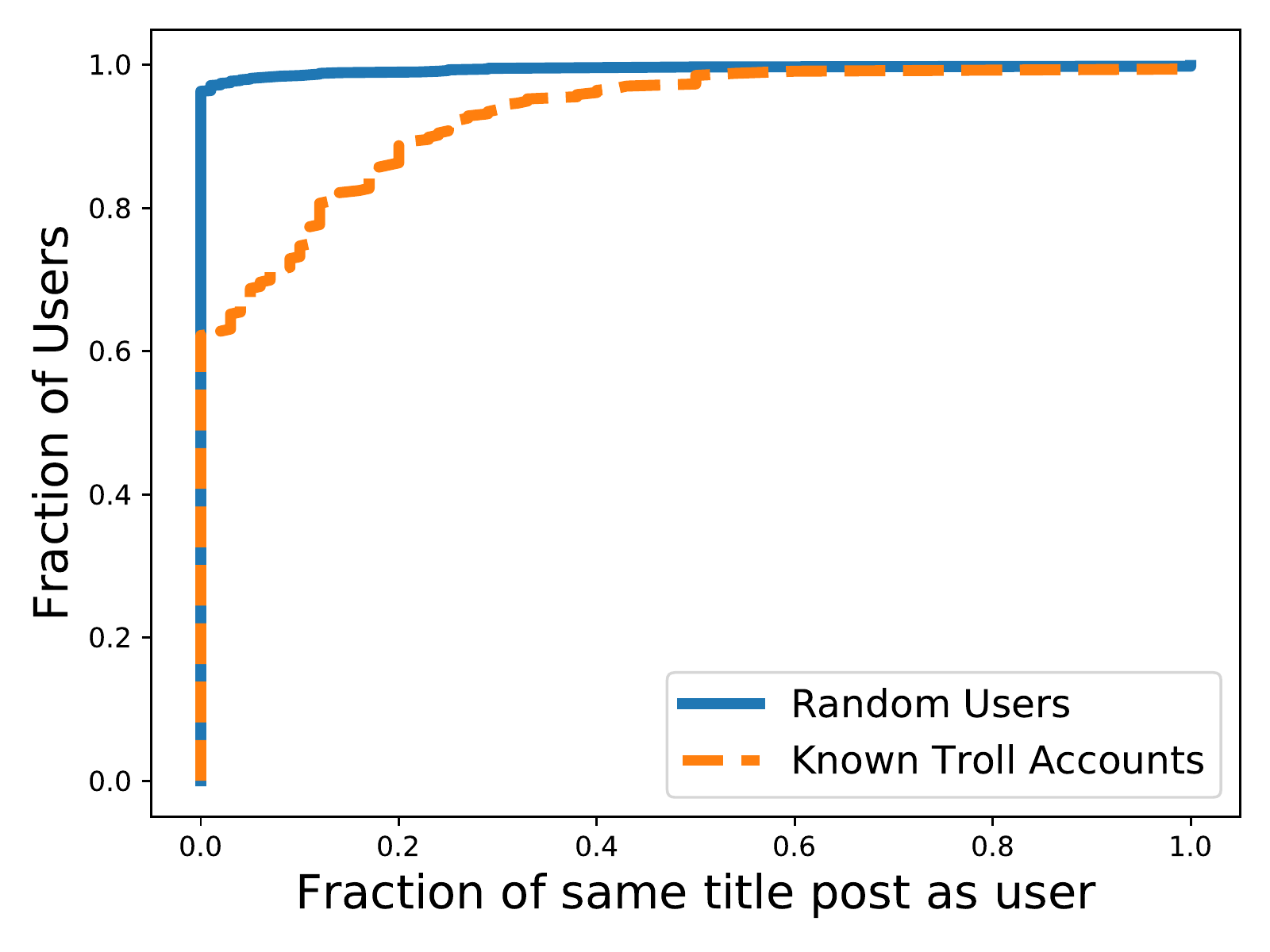}\label{fig:same_title}}
  \caption{Cumulative distribution functions (CDFs) of the fraction of comments in posts that users in the same class (a) started or (b) commented on, and (c) fraction of posts with the same title as posts made by another user in the same class.}
\label{fig}
\end{figure*}

\subsection{Dataset}\label{sec:dataset}
As ground truth, we use data released by Reddit on troll accounts active between 2015 and 2018~\cite{reddittrolls}, a timeline that includes the 2016 Brexit Referendum, the 2016 US Presidential Elections, and the 2018 US Midterm Elections.
The dataset includes \knowntrolls accounts, which generated 21,321 posts. %
We then collect a large set of Reddit accounts using the Pushshift public archives~\cite{baumgartner2020pushshift}.
These include all public posts made on Reddit in 2005--2020, accounting for 600M posts and 5B comments on 2.8M subreddits~\cite{baumgartner2020pushshift}.

\descr{Ethics.} Since we only use publicly available data and do not interact with human subjects, our work is not considered human subjects research by our IRB.
Also, we follow standard ethics guidelines: we only report aggregate data and do not deanonymize users.

\section{Characterizing Troll Activity}\label{sec:motiv}

As mentioned, our main intuition is that troll accounts show behavioral traits that are different enough from regular users to allow for accurate automated detection.
In particular, our hypothesis is that the interaction patterns shown by troll accounts controlled by the same actor will reveal patterns typical of loose coordination.

To investigate the viability of this approach, we analyze the activity of the \knowntrolls Russian-sponsored troll accounts described in Section~\ref{sec:dataset}.
As a baseline for comparison, we also extract all the posts and comments made by a set of random accounts from the Pushshift dataset. 
More precisely, we first identify the top 50 subreddits where the troll accounts were active; these include general audience communities like r/News and r/Politics as well as more specialized ones like r/Bitcoin.
We then extract \randomaccounts random accounts from these subreddits.

We analyze these two groups of accounts along three dimensions.
First, we want to investigate the assumption that troll accounts are more likely to comment on posts that were started by other troll accounts than random users are to comment on posts by other random users.
This has already been observed, at least anecdotally~\cite{mueller2019mueller}, as troll accounts might try to simulate legitimate discussion to push their disinformation narratives among regular users.
Figure~\ref{fig:comments_started} %
shows that troll accounts are indeed more likely to comment on posts started by other troll accounts than random users are on posts by another random user (2-sample KS statistic = 0.143, $p < 0.001$).

Second, we compare threads receiving comments from two or more troll accounts to those where more than one of our random users left a comment. 
The rationale is that troll accounts simulate exchanges of opinions %
aiming to polarize discussion and entice legitimate users to chime in~\cite{mueller2019mueller}.
Figure~\ref{fig:same_comments} confirms that troll accounts are more likely to comment on the same posts as other trolls than two random users are, on the same post (2-sample KS statistic = 0.132, $p < 0.001$).

Third, we look at posts created by troll accounts with the same title as posts by other troll accounts. %
The idea is not that they spam the same message multiple times (in fact, this would be trivial to detect) but that, when sharing a link to a Web page, the title of the Reddit post is set by default to that of the Web page.
Therefore, we expect that multiple accounts sharing the same news article (e.g., as part of a disinformation campaign) create posts with the same title.
Figure~\ref{fig:same_title} confirms that this is the case: troll accounts are significantly more likely to share two posts with the same title than random accounts are (2-sample KS statistic  0.346, $p < 0.001$).

Overall, this shows that troll accounts do behave differently than regular accounts on Reddit, and indicates that we can leverage behavioral features to automatically identify them.

\begin{figure*}[!t]
\centering
	\includegraphics[width=0.925\textwidth]{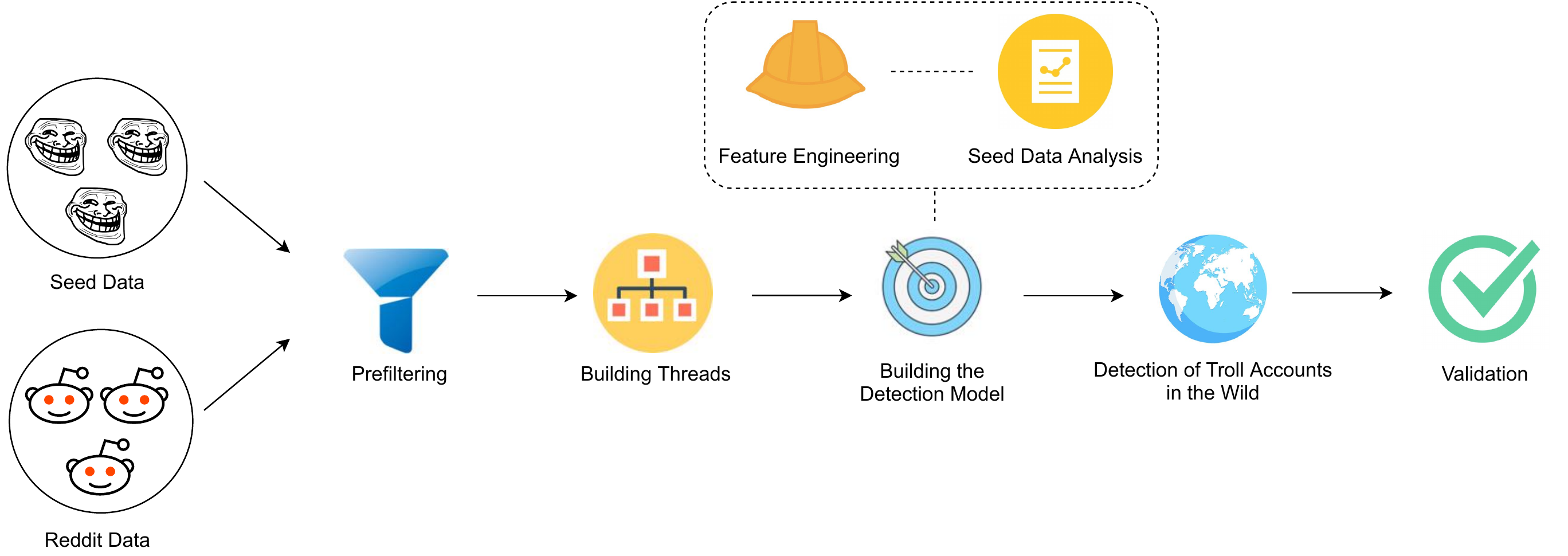}
	\caption{Overview of \system: two input streams are fed to the system: (1) a seed dataset of known troll accounts and (2) the entire Reddit data. \system filters a set of accounts from the Reddit dataset that have a high likelihood of being troll accounts, then, it  builds the thread structure of comments as seen on Reddit. Next, a detection model is built using pre-identified features and used to detect troll accounts in the wild. Finally, \system further analyzes the detected accounts to provide additional details about them.}
	\label{fig:systempipeline}
\end{figure*}

\section{Overview of \system}
\label{sec:methodology}

In this section, we present the analysis pipeline instantiated by \system;
it takes a seed dataset of known troll accounts and analyzes Reddit to find more accounts that behave in a similar fashion %
and are likely troll accounts.

\system operates in five stages:
1)~First, it identifies a set of suspicious accounts that have a higher likelihood to be troll accounts, compared to random Reddit accounts,
then, 2)~extracts all comments and submissions by those accounts %
and builds a thread structure from these posts.
3)~Next, \system trains a detection model based on a set of pre-identified features, and 4)~uses this model to detect troll accounts in the wild. 
5)~Finally, \system provides a number of additional analyses, giving moderators more details on their activity and allowing them to make informed decisions on whether the account needs to be suspended.

\subsection{Pre-Filtering}\label{sec:dataexpand}
In the first step, \system identifies a set of accounts that are likely to be troll accounts.
We do so to obtain a dataset of accounts that have similar posting habits as troll accounts and thus has a higher chance of containing potential troll accounts. 
Based on the observations highlighted in Section~\ref{sec:motiv}, \system considers a Reddit account as a potential troll account if they do any of the following:

\begin{enumerate}
\item \textit{Commenting on a troll account's submission}: As shown in Figure~\ref{fig:comments_started}, troll accounts are more likely to comment on a thread that was started by another troll account than by a random one.
As mentioned, this is done to simulate genuine interaction and lure unwitting users into the discussion. %

\item \textit{Making a submission with the same title as a troll account:} As evident from Figure~\ref{fig:same_title}, troll accounts are more likely to post submissions with the same title.
This is not necessarily due to troll accounts manually selecting the same title, but a side effect of how Reddit works: by default, when posting a URL, the submission's title is set as the title of the target page.

\end{enumerate}
	
\rev{Note that, unlike the features later used by \system for classification, these pre-filtering conditions are only determining whether a Reddit account has shown activity that may be indicative of it being a troll.
In later steps, \system analyzes, in detail, the activity of this candidate set of accounts, looking at how similar their activity is to known troll accounts, thus making a classification decision.
}

\subsection{Building Threads}
After \system identifies a set of suspicious accounts based on these two indicators, it proceeds to further process their data as collected from Pushshift.

As discussed, comment threads on Reddit are organized as trees.
This %
allows us to identify the specific comment a user replied to, and identify groups of users who have been conversing with each other.
However, the Pushshift data (which we use as our source) does not return these trees but provides comments in a flat structure.
In this step, \system parses these comments to build the comment tree, which is then used in the next steps.

\gs{Do we need this paragraph?}
More precisely, Pushshift stores each comment as a separate JSON object, where the \textit{link\_id} is the ID of the submission and the \textit{parent\_id} is the ID of its parent comment.
If the \textit{link\_id} and \textit{parent\_id} are the same, the comment is a top-level comment; i.e., it is a direct reply to the submission.
A comment can have any number of replies. 
To build the comment tree, \system extracts \textit{link\_id} of all comments made by the known troll accounts. 
Then, it uses the list of IDs to query the Pushshift data and finds all the comments on those submissions, not just the ones made by the known troll accounts.
It also uses a list of submission IDs of known troll accounts and finds all the comments which have a \textit{link\_id} from that list to find all comments on troll account submissions.
Once this data is retrieved, the \textit{link\_id}'s and \textit{parent\_id}'s are used to recreate the comment threads.

\subsection{Building the Detection Model}
\label{sec:model}
Next, \system builds a machine learning model to distinguish between legitimate and troll accounts.
To do so, we first manually select features based on our observations from Section~\ref{sec:motiv}.
We then train supervised learning models to perform detection.

\descr{Feature Engineering.}\label{sec:featureeng}
As discussed earlier, troll accounts exhibit recognizable behaviors, %
which we translate into nine features. %
In the following, we describe each feature and the reason why we select them. %

\begin{enumerate}
	\item {\em Total Comments:} From our analysis in Section~\ref{sec:motiv}, we find that troll accounts post 21 comments on average, compared to 300 for a random account. Therefore, we use the total number of comments made by an account as a feature.
	
	\item {\em Total Submissions:} Troll accounts make an average of 42 submissions, while a random account makes 32. Therefore, we use the total number of submissions made by the user as a feature.
	
	\item {\em Account Age:} Troll accounts are often created in waves, at or around the same time~\cite{zannettou2019disinformation}. Therefore, we select the time elapsed (in years) since the first submission or comment made by the user as a feature. 
	
	\item {\em Fraction of submissions with the same title as troll accounts:} Figure~\ref{fig:same_title} shows that troll accounts are much more likely to make a post with the same title as another troll account. 
	To include this information into our model, we calculate this feature as the fraction of submissions by the user that have the same title as a known troll account's submission.
	
	\item {\em Fraction of comments on submissions that troll accounts commented:} 
	Recall that troll accounts are more likely to comment on the same posts (see Figure~\ref{fig:same_comments}). 
	Therefore, we compute this feature as the fraction of comments by the user that are on a submission that a known troll account commented on.
	
	\item {\em Fraction of comments on submissions by troll accounts:} 
	Figure~\ref{fig:comments_started} shows that troll accounts are more likely to comment on posts started by other troll accounts. 
	To account for this, we add this feature as the fraction of comments by the user on a known troll account's submission. 
	
	\item {\em Fraction of direct replies on submissions by troll accounts:} 
	  Troll accounts have 0.5 direct replies on average on submissions by troll accounts, while a random account has none. Therefore, the fraction of direct comments (excluding comment threads) by the user on a known troll account's submission is used as a feature.
	
	\item {\em Fraction of comments that are a reply to a troll account's comment:} 
	  We find 49 instances of troll accounts replying to each other in comments, while random accounts never interact with each other.
	Therefore, we add this feature as the fraction of comments by the user that are a reply to a known troll account's comment.
	
	\item {\em Fraction of comments that are a reply to a troll account's comment in a troll account's submission:} 
	  We find 25 instances of troll accounts replying to each other in the comments under a submission made by another troll account, while random accounts never interact this way.
	To capture this information, we add this feature as the fraction of comments by the user that are a reply to a known troll account's comment on a known troll account's submission. 
	
\end{enumerate}

\descr{Building the model.}
Based on the set of features discussed above, \system trains a supervised model to distinguish between troll and legitimate Reddit accounts.
As we discuss in detail in Section~\ref{sec:validating}, we experiment with four classifiers: K-nearest neighbors~\cite{deng2016efficient}, Decision Tree~\cite{swain1977decision}, Support Vector Machine (SVM)~\cite{hearst1998support}, and Random Forests~\cite{breiman2001random}.

\subsection{Troll Detection} %
Once \system is trained, it can be used to detect troll accounts in the wild, as a classification task. %
That is, \system will return a set of detected troll accounts, which then go through further analysis and validation. %

\subsection{Validation}\label{sec:further_analysis}
After detecting accounts that are likely troll accounts, \system performs additional analysis to identify more indicators providing evidence that the accounts indeed belong to troll campaigns.
These checks serve to both validate our results %
and assist Reddit moderators by providing additional insights into the accounts.
\rev{To provide a comprehensive validation of its detection results, \system performs two types of analyses.
First, it looks at the detected accounts at an individual level, looking for indicators that they are likely trolls.
Second, it takes all detected accounts as a group, and looks at similarity in their collective activity with that of accounts in the seed set.
}
Note that this step takes into account elements that are not used as detection features by \system.

\descr{Account-level indicators.}
\rev{To further analyze detected troll accounts at an individual level \system looks at four aspects: 1) whether the account was deleted or suspended; 2) whether the account deleted any of their comments or submissions; 3) whether the account was created on the same day as one of the known troll accounts; and 4) whether the account posted a submission or a comment containing one of the top keywords used by the known trolls.
In the following, we discuss there four criteria in detail.}

\noindent\emph{Active Status.}
While they were not identified as state-sponsored troll accounts by Reddit, it is possible that the accounts identified by \system triggered other detection systems and were subsequently suspended, or that they were reported by Reddit users and later blocked by moderators. 
Therefore, \system checks whether the detected accounts have been banned or suspended by Reddit as an additional indicator of troll behavior.

\noindent\emph{Deleted Messages.}
Previous work~\cite{zannettou2019disinformation} shows evidence of troll accounts deleting their comments and submissions after the fact to avoid detection.
\system looks for evidence of accounts deleting their comments and submissions by comparing the data collected from the Pushshift API data with real-time data retrieved from the Reddit API.

\noindent\emph{Creation Date.}
Many troll accounts are created in waves, which means that they have the same creation date~\cite{zannettou2019disinformation}. 
To check for this, \system extracts the creation date of detected accounts from Reddit, and groups detected accounts together with known troll accounts based on this date.
Accounts that were created on the same day as known troll accounts have a much higher chance to be actual trolls. %

\noindent\emph{Topics Discussed.}
\rev{
As discussed in previous work~\cite{starbird2019disinformation,zannettou2019disinformation,zannettou2019let}, troll accounts push specific narratives and common talking points, which often reflect the geopolitical interests of the countries that control them. 
To analyze this aspect at an individual account granularity, we first identify the most important words shared by known troll accounts, and then check whether an account detected as a troll by \system posted about any of these words.
To do this, we calculate the TF-IDF (Term Frequency-Inverse Document Frequency) of the corpus of messages shared by known troll accounts~\cite{tfidf}. 
We then select the top 10 keywords identified by this approach as a proxy for the important narratives shared by known trolls, and check if a detected account included each of those keywords in any of their submissions or comments.
}

\descr{Group-level indicators.}
\rev{In addition to looking at accounts at an individual level, \system analyzes all detected accounts as a whole, to help identifying patterns of coordinated inauthentic activity.
In particular, we build language models on the comments posted by known and detected accounts as well as posting time patterns between the two sets of accounts.
}

\noindent\emph{Language Analysis.}
To further analyze the language used by detected troll accounts, \system builds language models based on word embeddings from the posts made by Reddit accounts, aiming to compare the language used by the detected troll accounts to that of known troll accounts and of undetected accounts.
This allows us to measure the similarity between the language used by known troll accounts, detected troll accounts, and other accounts, to investigate whether the detected troll accounts indeed use language that is closer to the known set of troll accounts.

\noindent\emph{Time Series Evaluation.}
Troll accounts often carry out their disinformation campaigns at specific points in time~\cite{zannettou2019let}, thus, it is likely that they will show similar activity patterns.
\system builds time series for known troll accounts, detected troll accounts, and non-troll accounts.
It then computes correlation and lag between the time series to confirm that detected accounts show higher coordination with known troll accounts as compared to non-troll accounts.

\section{Evaluation}
In this section, we present the results of our experiments running our Reddit dataset through \system.
\rev{We first discuss the results for each of \system' analysis steps, from pre-filtering to validation.}
\rev{We then present additional experiments to estimate \system's false negatives and to show that its approach can work on other influence campaigns, not only on the Russian sponsored one.
Finally, we report results on the run-time performance of \system.}

\subsection{Pre-filtering}

As discussed in Section~\ref{sec:dataexpand}, \system first identifies a set of suspicious accounts that present one of these traits: 1) posted the same submission title as troll accounts, or 2) commented on submissions made by troll accounts.
\system found 12,143 accounts that posted the same submission titles as troll accounts and 42,001 accounts that comment on submissions made by troll accounts. 
There is an intersection of 381 accounts between the two categories.
In total, this yields \expandeddataset accounts that are further analyzed. %

\subsection{Building threads}
We extract the comments and submissions of these accounts from the Reddit data published by Pushshift. 
The comments and submissions are used to calculate features to train the classifier.
In total, we collect 161,906,549 submissions and 938,852,501 comments made by the suspicious accounts.
Then, we build the thread structure for all submissions troll accounts commented on, resulting in 159,255 threads with an average depth of 2.69 and a median of 2. %

\subsection{Building the Detection Model}	
\label{sec:validating}

\begin{table}[t!]
	\begin{center}
		\setlength{\tabcolsep}{3pt}
		\small{
			\begin{tabular}{lrrrr} 
				\toprule
				{\bf Classifier} & {\bf Precision} & {\bf Recall} & {\bf Accuracy} & {\bf F1-Score} \\ %
				\midrule
				KNN	& 91.9\%	&	91.7\%	&	91.8\% & 91.8\% \\ %
				Linear SVM	&	95.7\%	&	95.5\%	&	95.5\%  & 95.6\% \\ %
				Decision Tree & 97.3\%    &  97.3\%     &  97.3\%     & 97.3\% \\ %
				\textbf{Random Forest}\hspace*{-0.25cm}	& \textbf{97.8\%}	&	\textbf{97.7\%}	&	\textbf{97.8\%}  & \textbf{\fscore\%} \\ %
				\bottomrule
			\end{tabular}
		}
	\end{center}
	\vspace{-0.2cm}
	\caption{Classification scores for troll account detection.}
	\label{classTable}
\end{table}

We extract a balanced dataset for training, with the set of \knowntrolls known troll accounts as the positive class and a random set of \knowntrolls accounts from the pre-filtered dataset as the negative class.
The reason we select accounts from the pre-filtered dataset instead of random Reddit accounts is to avoid over-fitting and to train a classifier geared to pick up subtle differences between the behavior of troll and non-troll accounts. 
Without doing so, \system would likely learn to flag any account that ever interacted with a known troll as malicious.
\sz{i am a bit confused. why the negative class is extracted from the prefiltered dataset? this is supposed to be a set of suspicious accounts no?}\gs{See if the explanation makes more sense now}
When selecting the random accounts for the negative class, we also ensure that these are not suspended by Reddit, to reduce the chances of them being troll accounts.

As discussed in Section~\ref{sec:model}, we experiment with four classifiers: KNN, Decision Tree, Linear SVM, and Random Forest.
To select the classifier best suited for the task, we perform 10-fold cross-validation.
We evaluate the performance of each classifier based on accuracy, precision, recall, and F1-score. 
Table~\ref{classTable} reports the average results using 10-fold cross-validation for each classifier. 
Although all classifiers perform well overall, Random Forest performs the best, achieving an F1-score of \fscore\%. 
Consequently, we use Random Forest for the detection model of \system, training on the whole training set of \knowntrolls troll and \knowntrolls random Reddit accounts.

\subsection{Detection in the Wild}

After training, we run \system on the entire dataset of \expandeddataset suspicious accounts to detect more troll accounts. 
This results in identifying \detectedtrolls accounts as trolls. 
In the next section, we provide further evidence that these accounts are likely to be troll accounts. %

\subsection{Validation --- Account-level Indicators}
\label{sec:validation-individual}
\rev{
As explained in Section~\ref{sec:further_analysis}, as a first step, \system checks each detected account individually for four indicators. 
}

\descr{Active Status.}
As discussed previously, an account's suspension is further evidence that the account is indeed a troll. 
To check whether an account exists on Reddit, we can look up reddit.com/u/$<$profilename$>$.json.
If the account was suspended, we get a 403 HTTP error; if it was deleted, the HTTP error code is 404. 
We find that 298 out of the \detectedtrolls accounts were either suspended or deleted.

\descr{Deleted Messages.}
We use PRAW: The Python Reddit API Wrapper~\cite{praw} to query all comments and submissions of detected trolls that are visible on their Reddit page as of April 14, 2021.
We then compare the number of comments and submissions for each detected troll account with those present in the data that we previously collected from the Pushshift API.
Our results show that 304 out of the \detectedtrolls detected trolls have deleted at least a comment or a submission, with 21 accounts having deleted all their comments and submissions.
It is important to note that PRAW only returns the last 1,000 comments/submissions, therefore we only count a deletion if PRAW returns less than 1,000 elements.
There are 14 accounts that hit the API limitation.
Also, we exclude the 298 deleted/suspended accounts because PRAW returns an error code for them.

\descr{Creation Date Analysis.}
We collect the Cake Day (or Account Creation Date) from the Reddit user's page of each of the known and detected troll accounts.
This excludes deleted and suspended accounts as their user page is not accessible on Reddit.
However, the user page of known troll accounts is still accessible, despite the suspension, as Reddit left them open for research purposes. 
We cluster the accounts by their creation dates and find that 66 out of the \detectedtrolls detected accounts belong to troll clusters making them highly suspicious.

\descr{Topic Discussed.}
\rev{ 
To identify relevant words discussed by the known trolls, we calculate the TF-IDF (Term Frequency-Inverse Document Frequency) of the corpus of submissions and comments that they posted~\cite{tfidf}. 
The TF is calculated on the known troll account dataset and the IDF on the entire dataset of \expandeddataset accounts.
Table~\ref{word_sim} reports the list of top 10 words shared by known trolls by this metric.
We then look at whether each of the \detectedtrolls detected accounts has posted a submission or comments containing one of those keywords.
We find that 359 of the detected accounts have at least one post containing one of the top 10 keywords shared by known trolls, indicating that they might be trying to push the same narratives as the seed set.
}

\descr{Summary.}
\rev{
Out of the \detectedtrolls accounts detected, 298 have been suspended/deleted, 304 deleted some of their comments/submissions, 66 were created on the same day as known troll accounts, and 359 posted a comment or submission containing one of top 10 keywords pushed by known troll accounts.
Overall, 824 accounts satisfied at least one of four conditions, accounting for 66\% of the \detectedtrolls detected accounts.
195 accounts satisfy two of the four conditions, 8 accounts satisfy three, and none of the accounts satisfy four.
}

\subsection{Validation --- Group-level Indicators}
\label{sec:validation-group}

\rev{
As discussed, in addition to looking at detected accounts in isolation, \system also analyzes the group of detected accounts collectively, to uncover additional insights on their coordination and provide further evidence that they are part of influence operations.
In the following, we first analyze the language used by the detected accounts compared to the known trolls; then, we look at the timing of their activity on Reddit.
}

\descr{Language Analysis.}
We use Natural Language Processing techniques to analyze the content of posts made by Reddit accounts, aiming to provide extra evidence that the detected accounts likely belong to influence campaigns.
This is particularly relevant, as \system is content-agnostic and does not look at the content posted by accounts. %
Therefore, finding language similarities at this stage of the analysis is a strong indicator that the detected accounts belong to the same disinformation campaigns as the set of known trolls.

To this end, we first train word2vec models on three corpora, each including submissions and comments posted by: 1) known troll accounts, 2) detected troll accounts, and 3) accounts from the set of \expandeddataset accounts that were not detected as trolls by \system.
For our word2vec models, we use Continuous Bag Of Words (CBOW), with a window size of 20, using the Python gensim library~\cite{gensim}.

\begin{table}[t!]
	\begin{center}
	\setlength{\tabcolsep}{3pt}
		\small
\begin{tabular}{lcccc}
\toprule
\textbf{Word} & {\bf Detected Trolls and}  & \textbf{Non-Trolls and} & {\bf Z-score} & {\bf P-Value} \\
& {\bf Known Trolls} & {\bf Known Trolls} \\ \midrule
people           & 0.53                                                                                       & 0.01                 & 7.55 & \textless .00001                                                                \\
money         & 0.35                                                                                        & 0.00				& 4.38 & \textless .00001                                                                                    \\
crypto        & 0.25                                                                                        & 0.01             & 2.12 & 0.03                                                                      \\
bitcoin       & 0.12                                                                                        & 0.01 &  1.27 & 0.20                                                                                   \\
country       & 0.12                                                                                        & 0.02  & 1.58 &  0.11                                                                                \\
police        & 0.12                                                                                        & 0.00   & 1.53  & 0.13                                                                               \\
black         & 0.11                                                                                        & 0.00    & 1.63 & 0.10                                                                                \\
news          & 0.08                                                                                        & 0.00  & 1.57            & 0.12                                                                       \\
cop           & 0.08                                                                                        & 0.01    & 0.82    & 0.41                                                                          \\
trump         & 0.08                                                                                        & 0.00     & 1.48 & 0.14                                                                               \\ \bottomrule
\end{tabular}%
		
	\end{center}
\vspace{-0.25cm}
	\caption{For each keyword, we obtain a vector of top-100 similar words from the word embeddings. The cosine similarity between the vectors of detected troll accounts and known troll accounts is higher for each keyword. \rev{The z-score and the corresponding p-value is also given. }}
	\label{word_sim}
\end{table}

\rev{In the previous section we discussed how we identify a set of the most relevant words shared by known troll accounts for further validation.
At this step we use this list of relevant words to further study the use of language by detected troll accounts.
}
For each of these words, we compute the similarity between the trained word embeddings by using the method from Vivek et al.~\cite{kulkarni2014statistically}. 
Since the word embeddings are trained on different corpora, they have unique vector spaces which cannot be directly compared.
Therefore, for each keyword, we extract the Top-100 most similar words to it and represent them as a vector calculated from the word embeddings.

Next, we compute the cosine similarity between the vectors of 1)~detected troll and known troll accounts and 2)~non-troll and known troll accounts.
Table~\ref{word_sim} shows the language similarity for the top keywords calculated in the previous section. %
As it can be seen, for all words, detected troll accounts have a higher similarity than undetected accounts when compared to the language used by known trolls.
\rev{We also perform a z-test to show that known trolls use similar language as detected trolls but their use of language is different from non-trolls.
To calculate the z-score, we use the cosine similarities from each row of Table~\ref{word_sim} as proportion, where the population size is the number of messages containing a certain word.
Our results show that for three keywords (i.e., ``people,'' ``money,'' and ``crypto'') the differences in language show statistically significant differences at the $p < 0.05$ level.
In the remaining cases, although the cosine similarity between detected trolls and known trolls is higher than the one with undetected accounts, the test is inconclusive, likely due to the limited sample size.}

\begin{figure*}[t]
	\centering
	\subfigure[\textit{Known trolls}. Larger nodes represent words common with (b).]{\includegraphics[width=0.3\textwidth]{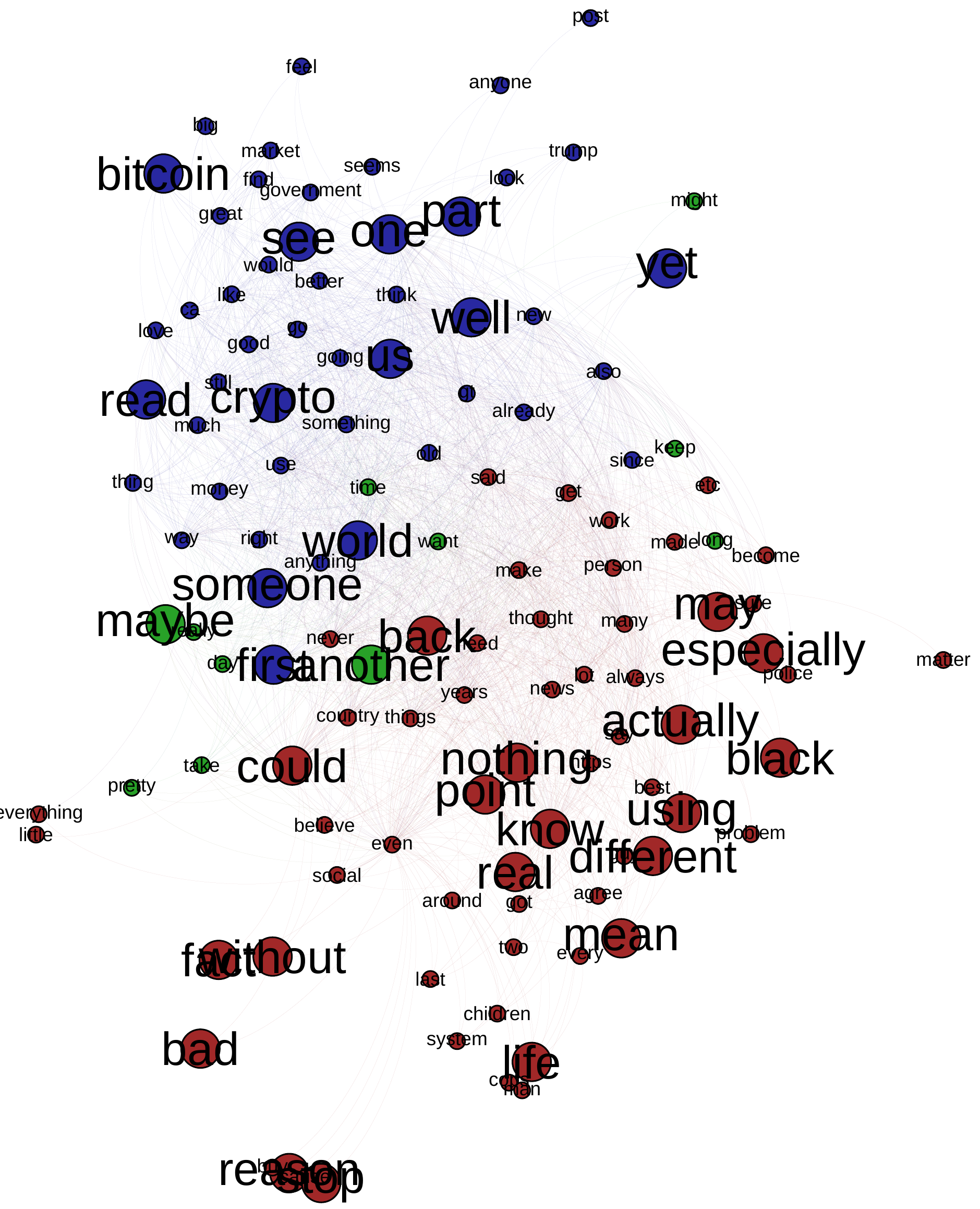}\label{fig:crypto1}}
	\hfill
	\subfigure[\textit{Detected trolls}. Larger nodes represent words common with (a).]
{\includegraphics[width=0.3\textwidth]{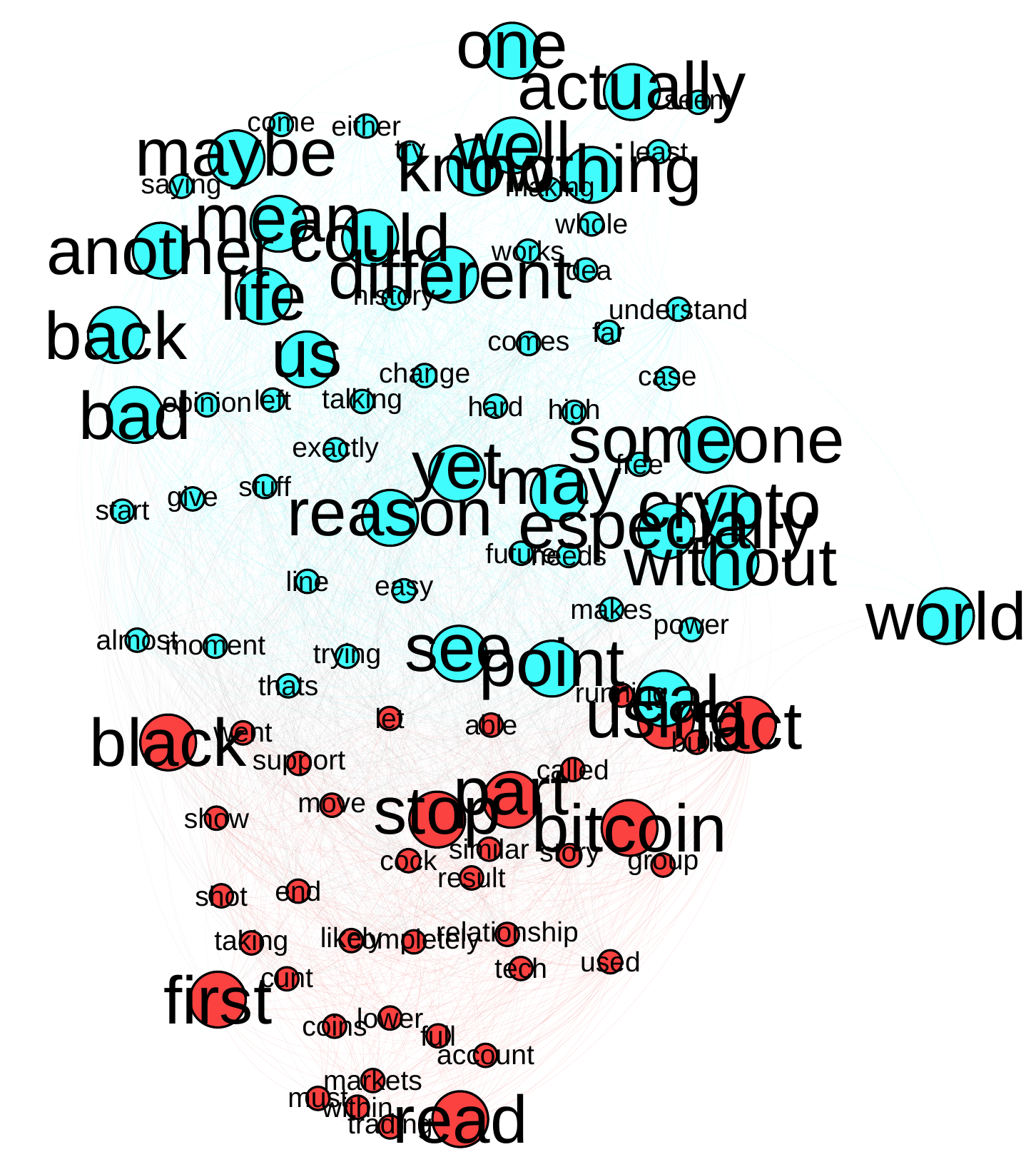}\label{fig:crypto2}}
	\hfill
	\subfigure[\textit{Non-trolls}. Larger nodes (``crypto" and ``bitcoin") are common with (a).]{\includegraphics[width=0.38\textwidth]{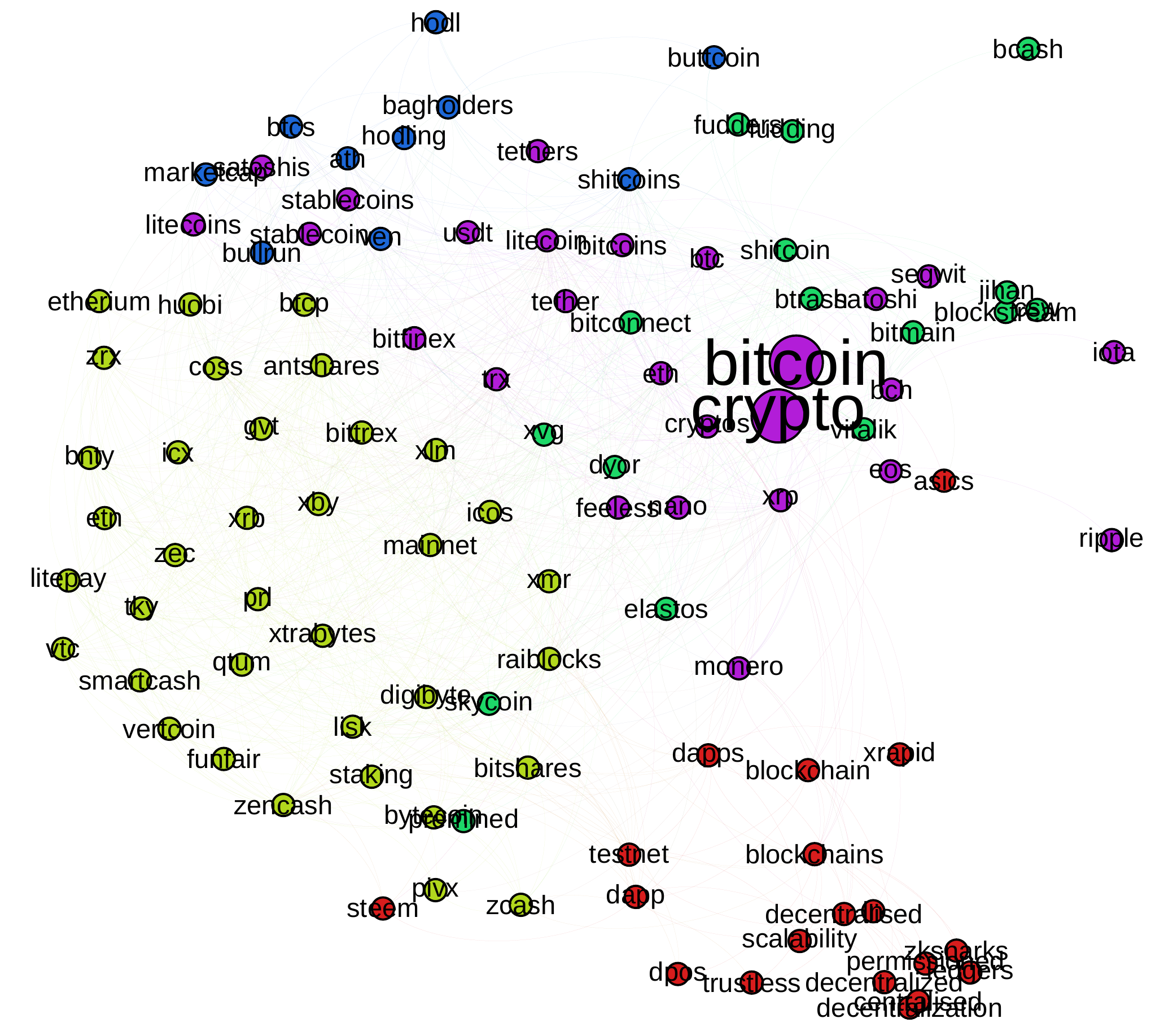}\label{fig:crypto3}}
\medskip
	\caption{A visualization of language usage in relation to the keyword ``crypto'' where nodes from the same community (detected using the Louvain community detection method~\cite{Blondel2008}) are depicted with the same color. It is evident that known trolls and trolls detected by \system have more words in common than known trolls and non-trolls. }
	\label{fig:word2vec_crypto}
\end{figure*}

\begin{comment}
\begin{figure*}[p]
	\centering
	\subfigure[\textit{Known trolls}. Larger nodes represent words common with (b).]{\includegraphics[width=0.325\textwidth]{Figures/black_trolls}\label{fig:black1}}
	\subfigure[\textit{Detected trolls}. Larger nodes represent words common with (a).]{\includegraphics[width=0.325\textwidth]{Figures/black_detected_trolls}\label{fig:black2}}
	\subfigure[\textit{Non-trolls}. Larger node (i.e. ``black") is common with (a).]{\includegraphics[width=0.325\textwidth]{Figures/black_non_detected}\label{fig:black3}}
\medskip
	\caption{A visualization of language usage in relation to the keyword ``black'' where nodes from the same community (detected using the Louvain community detection method~\cite{Blondel2008}) are depicted with the same color. It is evident that known trolls and trolls detected by \system have more words in common than known trolls and non-trolls. }
	\label{fig:word2vec_black}
\end{figure*}
\end{comment}

\descr{Deep-Diving on Keyword ``Crypto''.} 
To better illustrate the difference and similarity in language used by the different types of accounts we focus on the language surrounding the keyword ``crypto,'' since the z-score returned statistically significant results for it and the topic of cryptocurrencies is of great interest to the computer security communities and understanding how state-sponsored operation attempt to influence the field is an unstudied area.
The word ``crypto'' appears in 817 comments made by detected troll accounts and 132 comments by known troll accounts.
Also, there are 50 detected troll accounts and 16 known troll accounts with comments containing the word ``crypto.''

\descr{\em Visualization.} %
To visualize the language used in relation to our keywords, we follow the methodology proposed by Zannettou et al.~\cite{zannettou2019quantitative}.
Figure~\ref{fig:word2vec_crypto} present the graphs calculated from the word ``crypto.''
Nodes are words and are connected by an edge if the cosine similarity of their embedding vectors is above a given threshold.
The threshold for known troll and detected troll accounts is set to 0.9, whereas for the undetected accounts, the threshold is set to 0.68.
These thresholds are selected to keep approximately 100 nodes in each graph.
We chose 100 as the number of nodes to: 1) ensure consistency with the word embedding analysis where we compared vectors of Top-100 most similar words, and 2) have a reasonable number of nodes for visualization.  \sz{is there an intuition behind this?} \hs{added. does it look ok?}
The graph is built from the trained word2vec model and only nodes within two hops from the keyword are included.

To visualize the graphs, we perform a number of steps.
First, we construct a weighted graph using the ForceAtlas2 layout algorithm~\cite{gephi} where words with higher cosine similarities are laid out closer in the graph space.
Figure~\ref{fig:crypto1} shows the word embedding graph for troll accounts and the words with a larger font are common with detected troll accounts, Figure~\ref{fig:crypto2} those of detected troll accounts and the highlighted words are common with known troll accounts, while Figure~\ref{fig:crypto3} shows the word embeddings for undetected accounts and the highlighted words are common with known troll accounts.

\descr{\em Language Used.} The graphs indicate that the language used by detected troll accounts is indeed closer to that used by known troll accounts.
For ``crypto,'' the graph for the detected troll accounts (Figure\ref{fig:crypto2}) has 33 words in common with known troll accounts whereas the non-troll accounts one (Figure~\ref{fig:crypto3}) only has two (i.e. bitcoin and crypto). 

As an additional indicator, we run the Louvain community detection
algorithm~\cite{Blondel2008} on the graph, to identify ``communities'' of similar words.
Words belonging to the same community are depicted with the same color.
From Figure~\ref{fig:word2vec_crypto}, we observe that non-troll accounts discuss cryptocurrencies in general, covering a wide variety of coins.
On the other hand, known troll and detected troll accounts talk about bitcoin in particular and use more informal language.

\descr{\em Comment Examples.} To further illustrate the differences in language and topics covered by known, detected troll, and non-detected accounts, we discuss a few, manually selected, comments containing the word ``crypto'' by each class of accounts.
First, we look at three comments made by known troll accounts containing the word ``crypto.''
\sz{how are these 3 comments selected?}\hs{manually. should i add?}
\gs{added earlier}

\begin{mdframed}[style=exampledefault, backgroundcolor=vlightgray]\small
COMMENT 1: 
All my family members are trading crypto 

\noindent COMMENT 2:
I feel like the reporter had an anti crypto vibe to her.

\noindent COMMENT 3:
Crypto is down another 5\% since this news broke. Fuck this gay Earth.
\end{mdframed}

All comments take a pro-crypto stance and advocate the trading of cryptocurrencies. %
They also seem invested in cryptocurrencies and discredit a reporter if they are critical of cryptocurrencies (in Comment 2) or show anger if their price drops (in Comment 3).

The following comments are made by detected troll accounts and similar to the known trolls, they take a strong pro-crypto stance.

\begin{mdframed}[style=exampledefault, backgroundcolor=vlightgray]\small
COMMENT 1: 
No we need to destroy them.
no one bashes crypto currencys.

\noindent COMMENT 2:
I'm just living in Crypto 24/7. Everything is all-right ...
\end{mdframed}

Finally, the comments shown below are made by accounts that were not detected as troll accounts by \system. Contrary to the troll and detected accounts, they express frustration towards cryptocurrencies and an anti-crypto stance.

\begin{mdframed}[style=exampledefault, backgroundcolor=vlightgray]\small
COMMENT 1: 
God am I sick of all these small little start up based crypto coins. 

\noindent COMMENT 2:
Just be careful, no guarantees in crypto!

\noindent COMMENT 3:
All crypto currencies will go to zero eventually.
\end{mdframed}

\begin{figure*}[t]
	\centering
	\subfigure[Comments]{\includegraphics[width=0.485\textwidth]{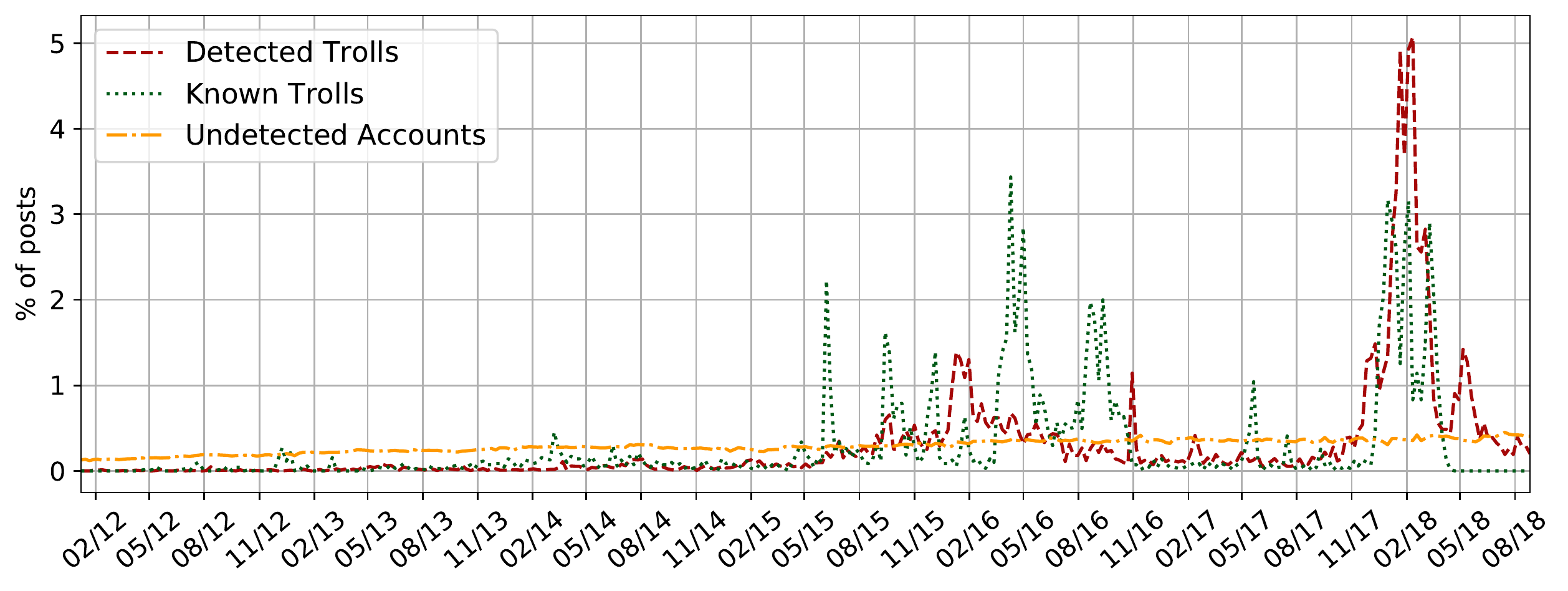}\label{fig:timeseres_com}}
	\subfigure[Submissions]{
	\includegraphics[width=0.485\textwidth]{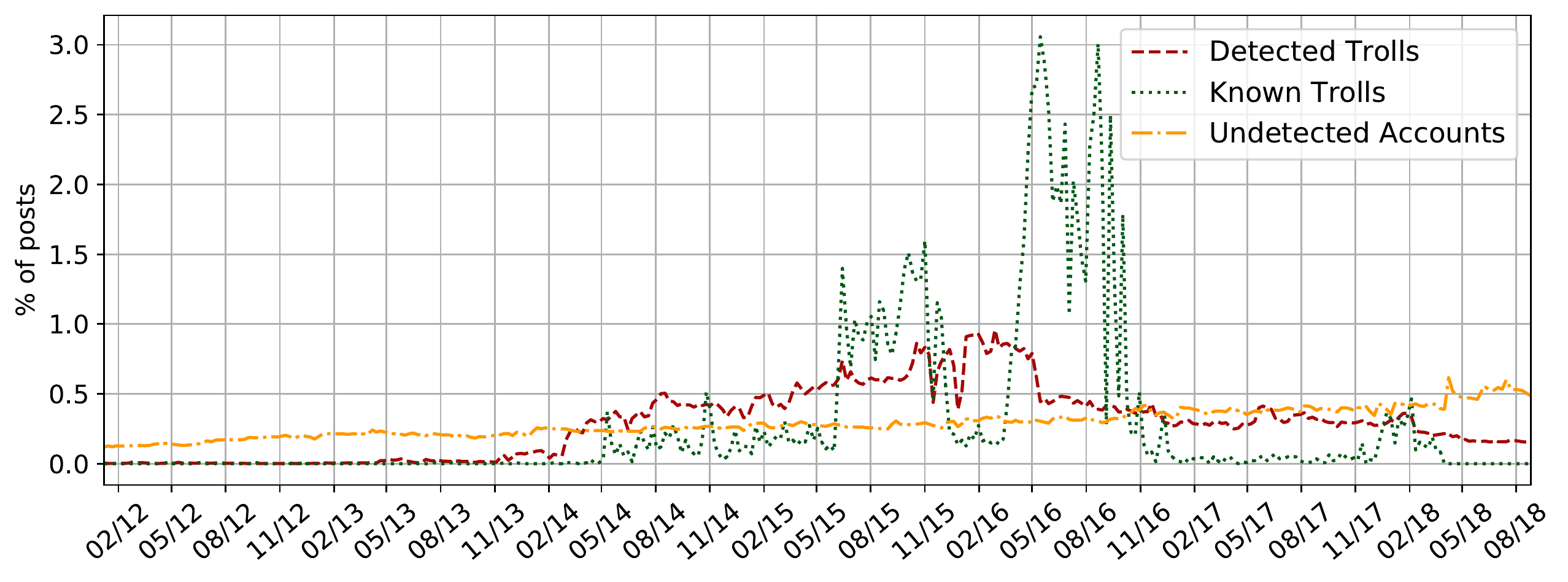}\label{fig:timeseres_sub}}
	\caption{Time Series of Comments and Submissions.}
\end{figure*}

\descr{Time Series Evaluation.}
Another crucial verification step is a time series analysis of the activity of accounts.
Our intuition is that troll accounts will show a time synchronization that is not evident for regular accounts.
To confirm this, plotted the time series of known troll accounts, detected troll, and accounts labeled as non-troll accounts by \system.

Figure~\ref{fig:timeseres_com} shows the time series plot for the comments. 
The activity of detected accounts is much closer to known troll accounts as compared to the non-troll accounts, especially the peaks in 2018.
Figure~\ref{fig:timeseres_sub} shows the time series plot for submissions.
Similar to the comments activity, detected troll and known troll accounts show a coordination pattern.

Next, we compute the Pearson correlation and lag for both time series to check the degree of similarity in the activities.
For submissions, the correlation between detected troll and known troll accounts is 0.5, compared to the 0.068 between undetected accounts and known troll accounts.
Similarly, for the first group, the lag is -23 and it is -99 for the other.
In the case of comments, the correlation for the first group is 0.553 and 0.334 for the negative group.
The lag is 93 for the first group and 173 for the other, showing clearly that the detected troll accounts are much more similar to known troll accounts than undetected accounts are.

\newrev{
To improve our validation, we repeat this experiment only considering the 424 accounts for which no individual account indicator was satisfied (in Section~\ref{sec:validation-individual}) and the same findings still hold (0.49 correlation for submissions and 0.54 for comments, with the group lag remaining unchanged).
This indicates that, although those accounts could not be confirmed individually, they present activity patterns as a group that are close to the known troll accounts.}

\subsection{Summary}
When looking at the set of detected accounts as a group, we find that detected troll accounts push similar narratives, use similar language to known troll accounts, and show a higher degree of synchronization compared to accounts that are not detected by \system.

\subsection{Estimating False Negative Rate}
\label{sec:false-negatives}

\newrev{
To estimate \system's false negatives, we perform a qualitative analysis where we randomly pick undetected Reddit accounts and check for signs that might be indicative of them being trolls.
We manually annotate a set of 20 random accounts from the set labeled as non-trolls by \system.
Two authors of this paper independently assessed the 20 accounts looking for inflammatory, insincere, digressive, extraneous, or off-topic messages.
Note that, while posting this kind of messages is not an ultimate indicator that an account is a troll, this allows us to establish an upper bound for \system's false negative rate.
Annotator 1 labeled 2 accounts as trolls whereas Annotator 2 labeled 1 account as a troll.
The account labeled as troll by Annotator 2 was also labeled as troll by Annotator 1, yielding an inter-coder agreement of 95\% and Cohen's Kappa score of 0.64 (i.e., {\em substantial} agreement).
Therefore, with 2 out of the 20 accounts labeled as trolls, we estimate that the false negative rate of \system is 10\%.}

\subsection{Measuring the Reach of Troll Accounts}
\label{sec:impact}

\rev{
An important question when studying troll operations relates to the impact they have on the platforms that they are active on, and on their users.
To this end, we now look at whether the submissions and comments by troll accounts receive more engagement than those by undetected accounts in our dataset.
Ideally, we would like to measure how many Reddit users saw a certain post, but unfortunately our dataset does not contain that information.
Instead, we use the {\em score}, namely, the number of upvotes minus the number of downvotes, as a proxy.}

\rev{We find that the troll accounts detected by \system have made 26,984 comments with a total score of 153,839. 
The average score per comment is 5.7, which is higher than the one for comments posted by the other accounts in our dataset i.e., 4.8.
This suggests that troll accounts receive more engagement than regular accounts on Reddit, opening up interesting research questions to be further investigated in future work.
}

\subsection{Testing \system on Another Campaign}
\label{sec:uae}
Our experiments show that \system can effectively identify troll accounts that belong to a specific Russian influence campaign identified by Reddit.
Next, we set out to verify whether or not \system can operate on other campaigns as well.
To this end, we run the entire pipeline on an UAE-sponsored influence operation, composed of 28 accounts active across several subreddits.
The list of 28 accounts was compiled by a journalist covering disinformation and was shared with the authors of this paper.
We download their comments and submissions using Pushshift API and extract features to train \system.
We also download the data for all 42,738 accounts that commented under their submissions and 445 that posted the same titles as these trolls.
We then extract a balanced dataset for training, with the set of 28 known troll accounts as the positive class and a random set of 200 accounts from the collected dataset as the negative class.

A 10-fold cross-validation using a Random Forest classifier achieves 99.3\% accuracy, 99.6\% precision score, and 98.3\% recall.
We then used the model trained on this ground truth to detect additional troll accounts belonging to the UAE campaign, and found 13 new accounts (out of the 43,183 accounts that passed the pre-filtering stage). Of these, 12 satisfy one or more of our individual account indicators.

Albeit preliminary, these results indicate that \system can learn the typical traits of state-sponsored disinformation campaigns on Reddit, and could be used to protect Reddit against emerging campaigns.

\subsection{Run-Time Performance} 
\label{sec:performance}
Last but not least, we discuss the scalability of running \system on a large social network platform.
\newrev{
We tested \system on a server with two 12-core Intel(R) Xeon(R) Gold 6126 @ 2.60GHz CPUs and 768GB ram.
When testing the runtime performance of \system in completing the various steps of the analysis pipeline, we found that while most steps are completed quickly, retrieving data from the Pushshift API is a significant bottleneck.
It currently takes \system, on average, 226 seconds to retrieve data for an account from the Pushshift API, 0.15 seconds to extract the features for our classifier, and 0.005 seconds to perform the actual classification. 
Taking aside data extraction, it takes \system 2.5 seconds to train our model.
We expect the data retrieval time to go down significantly when deployed in production by a large company with a powerful infrastructure such as Reddit.
Since performing detection for an account can be done independently from other detection tasks, this operation could be parallelized for as many accounts as needed (e.g., all the accounts that interacted with a known troll in a given day). 
We envision a batch deployment in the wild, where the troll detection algorithm is run at fixed intervals, for example once a day, in a similar fashion as proposed by state-of-the-art social network abuse systems~\cite{cao2014uncovering,stringhini2015evilcohort}.
}

\section{Case Studies}

\begin{figure}[t]
	\centering    
	\includegraphics[width=0.495\textwidth]{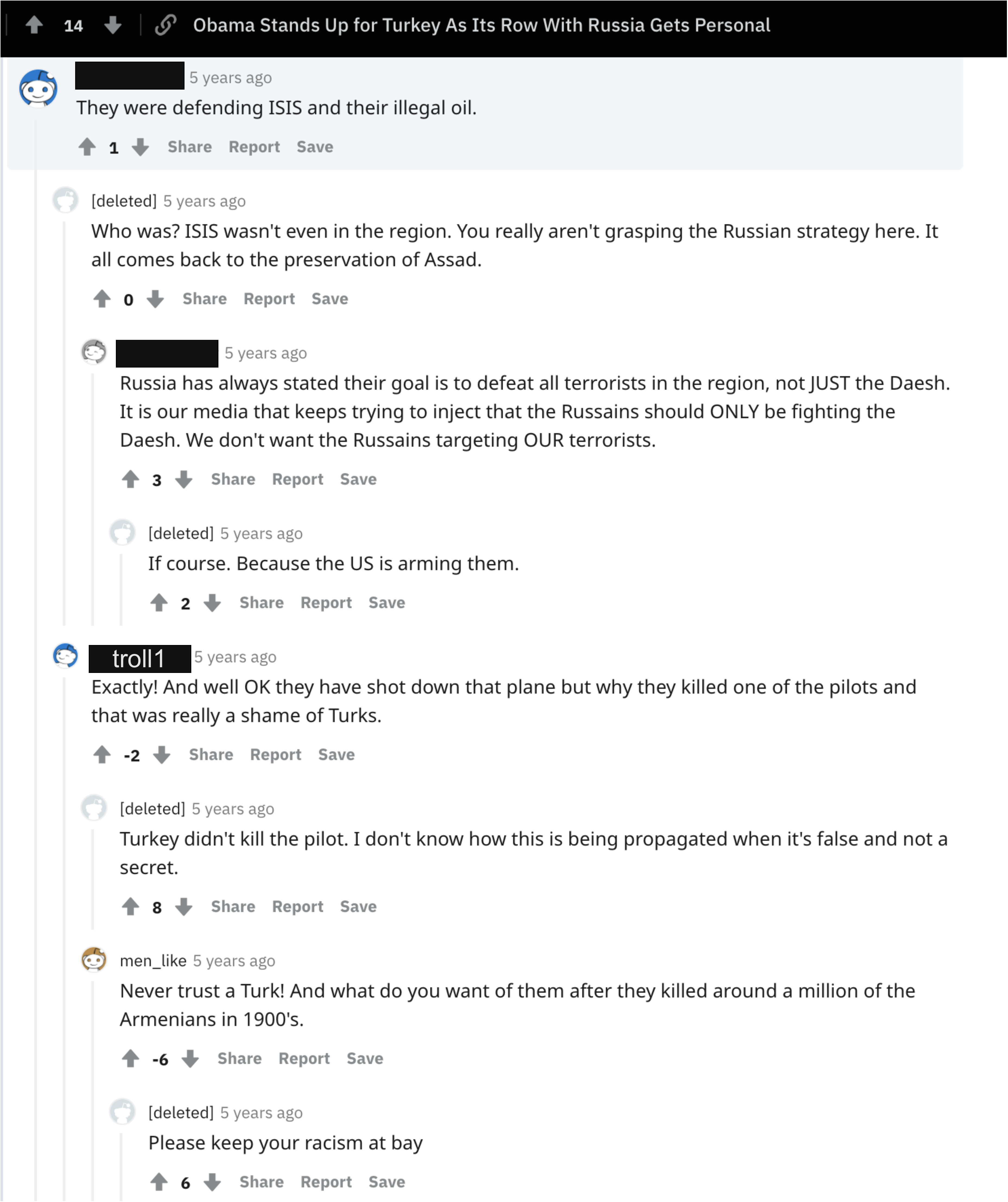}
	\caption{An example of manufactured conflict between known trolls and accounts detected by \system.}
	\label{fig:casestudy_collaboration}
\end{figure}

In this section, we present and discuss two case studies selected from the accounts identified by \system as troll accounts.
We do so to shed light on the modus operandi of the troll accounts, both with respect to spreading disinformation/polarizing online discussion and pretending to be real users by posting harmless content. 

\descr{Case Study 1: Manufactured conflict.}
Former operatives of Russian troll operations have explained that disinformation actors often worked in groups to polarize online discussions~\cite{rfe}.
For instance, they would have an account post a message, and other accounts vehemently disagree with it, aiming to attract real users and further polarize the discussion on that subreddit.

In our work, we observed similar instances of trolls ``teaming up'' and posting in groups of 2-3 accounts on the same submission, often replying to each other.
Figure~\ref{fig:casestudy_collaboration} shows one such case of collaboration on the subreddit \texttt{r/worldnews}.
This is a snippet taken from a 5-year old submission that was discussing then-President Obama siding with Turkey during tensions with Russia.
There are five accounts participating in this thread, two of which are of interest to us: \texttt{men\_like}, a known troll account identified by Reddit and an alleged troll account detected by \system, which we will refer to as \texttt{troll1} for privacy reasons.
There is also a third account that was deleted by its owner (marked as \texttt{[deleted]} in Figure~\ref{fig:casestudy_collaboration}).
In the comment thread, \texttt{troll1} first mentions an incident to shame the Turkish military.
Then, \texttt{men\_like} takes it further by suggesting that Turkish people are not to be trusted at all.
Finally, the now-deleted account accuses \texttt{men\_like} of racism.

This is a textbook example of a manufactured controversy on Reddit, designed to push a certain narrative with the goal of influencing the real users on that Subreddit.

\begin{figure*}[t]
	\centering
	\subfigure[]{\includegraphics[width=0.32\textwidth]{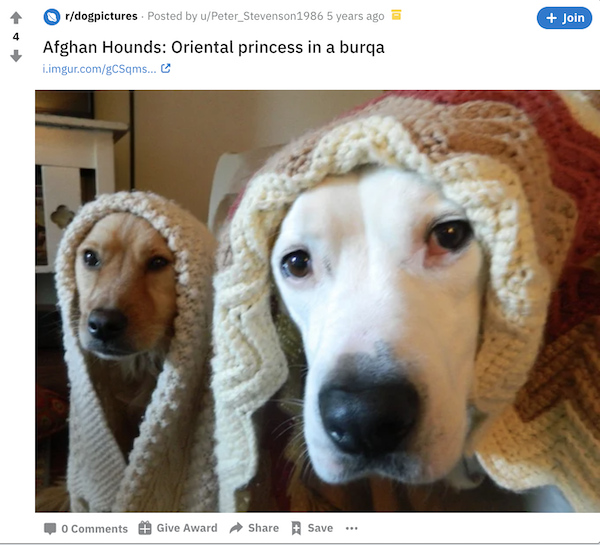}\label{fig:dog1}}
	\subfigure[]{\includegraphics[width=0.32\textwidth]{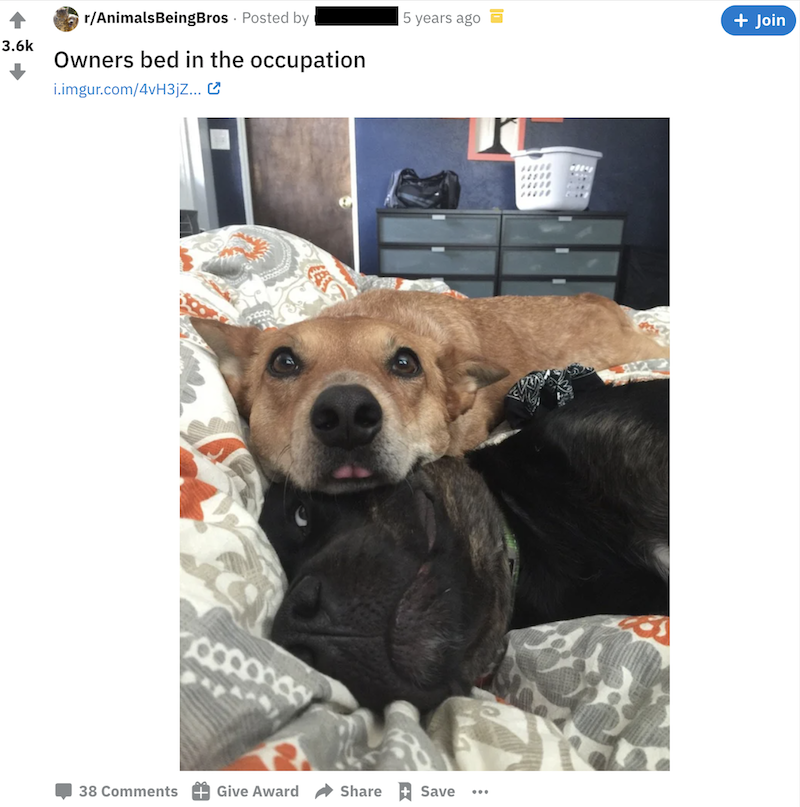}\label{fig:dog2}}
	\subfigure[]{\includegraphics[width=0.32\textwidth]{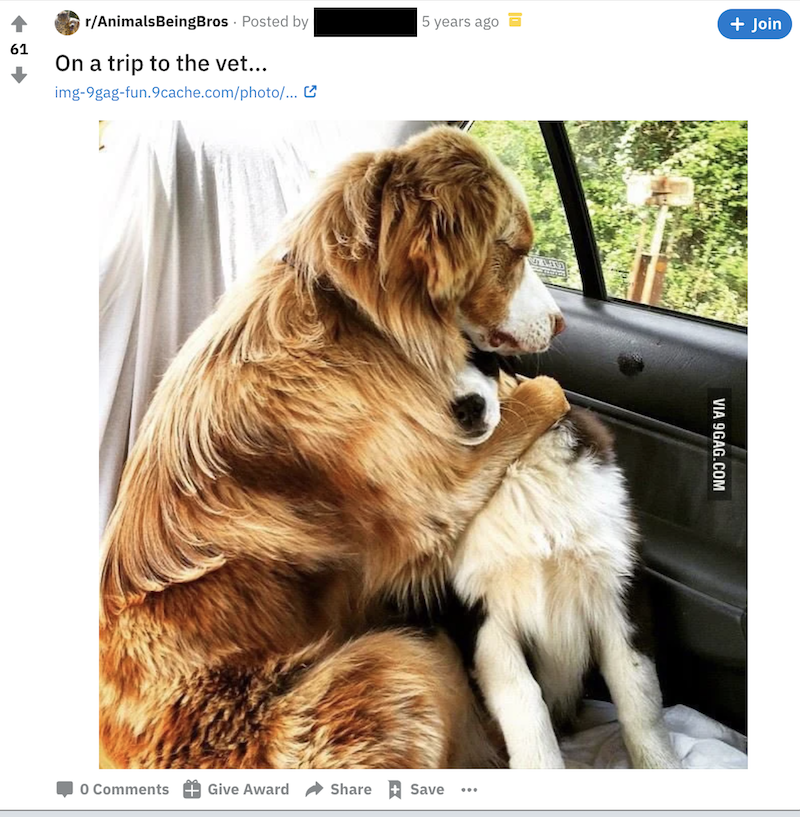}\label{fig:dog3}}
	\caption{The figure shows the similarity of posts made by known trolls and  accounts detected by \system. The left-most post is made by a known-troll and the other two are from accounts detected by \system.}
	\label{fig:casestudy_same_content}
\end{figure*}

\descr{Case Study 2: Simulating Legitimate Activity.} %
Another goal of the trolls is to appear legitimate to other users, as well as to Reddit itself, raising suspicion.
To this end, it makes sense for troll accounts to post content that is unrelated to their ``primary purpose.''
In our dataset, we find similarities in the type of ``benign'' content being posted by several accounts.
An example is presented in Figure~\ref{fig:casestudy_same_content}: the left-most picture is posted by one of the known troll account identified by Reddit, while the other two by two troll accounts detected by \system.
These three accounts post amusing pictures of dogs to look more ``legitimate'' and blend into the Reddit community.
Similarly, we find 36 accounts that participate in the \texttt{r/Jokes} subreddit and post funny memes.
The post of general purpose and seemingly irrelevant content/images is likely an attempt by the trolls to accumulate karma score on their account and pose as legitimate users, hence decreasing the likelihood of their detection.

\section{Discussion}
\sz{we can say somewhere (probably in discussion) that we show interactions are important and that Reddit has a richer set of interactions (e.g., who votes what) and this might be helpful in identifying trolls.}
In this section, we discuss important implications of our results for social media platforms. 
We also outline the ways in which attackers can leverage this information to their benefit. 
Finally, we highlight some limitations of our study along with possible future directions. 

\subsection{Implications for Social Media Platforms}

\jbnote{begin Jeremy wrods}
Disinformation on social media has become one of the most pressing issues in modern society, and is at the forefront of trust and safety initiatives across essentially all popular platforms\jbnote{should we cite like, twitter, facebook, reddit, discord, instagram, etc. here or just let reader not be stupid?}.
While related to the well-explored area of social bots, troll accounts represent a meaningfully different attack vector that must be addressed.

The key insight from our study is that \emph{loose} coordination via direct interaction (e.g., commenting on each other's posts) and focused narrative pushing (e.g., posting the same submission across different troll accounts) are core features of troll accounts.
Our study provides an automated mechanism for detecting troll accounts used in disinformation campaigns, and can serve as a blueprint for practitioners to build production detection systems.
If anything, our results indicate a \emph{lower bound} on the efficacy of automated troll detection systems.
Reddit themselves have a \emph{much} richer set of interactions than is publicly available (e.g., who upvotes what content/posts) and can thus model troll interactions at a higher resolution than we can.

\jbnote{end jeremy words}

\subsection{Resilience to Evasion}

\system's detection is designed to embody the key behavior of troll accounts and their need to coordinate to effectively spread disinformation narratives.
As with any machine-learning powered detection system, however, \system could be evaded by miscreants once they get to know how the model operates.
However, we argue that, to do so, attackers would need to significantly change their modus operandi, and that these adjustments would make their attacks more similar to traditional automated fake activity on social media, which can be detected by existing research.

For example, attackers could attempt to evade detection by posting an overwhelming amount of unrelated comments on legitimate threads.
While this might successfully evade \system, it might make the accounts stand out to existing detection systems that aim to identify bot-like and spam activity, defeating the purpose~\cite{benevenuto2010detecting,cao2014uncovering,davis2016bot,stringhini2010detecting,xudeep}. 

An alternative evasion strategy could be to have each troll account post less or interact with a smaller number of troll accounts or make fewer posts to avoid raising suspicion.
If miscreants adopted this solution, they would need to create a larger number of accounts to keep the same level of engagement, and this could be detected by other approaches that detect mass-created fake accounts~\cite{stringhini2015evilcohort,yuan2019detecting}.

Finally, malevolent actors might have their troll accounts never interact with each other.
However, this would make their operation much less effective, since a key part of their activity is based on creating conflict around sensitive topics by having these accounts talk to each other, as shown in this paper.

\subsection{Limitations}

Naturally, our work is not free from limitations.
First of all, \system requires a set of known troll accounts to bootstrap its capabilities.
This means that our approach cannot detect new and emerging troll campaigns that have not been observed before.
\newrev{The lack of ground truth has also limited the set of experiments that we could reliably run in this paper, forcing us to only work with a single state-sponsored campaign.
Techniques to establish rigorous ground truth for these problems are desperately needed by the research community, and could foster more research in this space.
}

Another limitation is our best-effort selection of legitimate accounts to train our classifier (see Section~\ref{sec:dataexpand}): that is, we cannot be 100\% sure that the set we selected does not contain any trolls.
Finally, while we provide evidence that the \detectedtrolls detected accounts behave like troll accounts, we do not have definite proof of that; however, we are in contact with Reddit to obtain further details/confirmation from them.

\subsection{Future Work}

\system presents a first-of-its-kind approach to improve defenses against troll accounts posting disinformation.
We envision several ways in which this work can be improved in future work.
First, \system requires a seed dataset of known troll accounts to be trained on; an interesting line of research would be to investigate behavioral features that are independent of the specific campaign that trolls belong to, allowing to generalize \system to previously unseen campaigns.

Second, we were limited in the number of influence campaigns that we could study due to the absence of reliable ground truth, and only focused on an influence campaign carried out by Russian-sponsored accounts and on a very small campaign carried out by UAE-sponsored accounts.
If further ground truth became available, future work should investigate whether \system can generalize to additional influence campaigns.

Additionally, one could look at the narratives being pushed by trolls. 
An interesting area of research is to analyze the strategies employed by these accounts and how they interact to spread a particular narrative. 
Further analyzing this behavior and how it affects legitimate social media users is critical to better understand the disinformation landscape.

\section{Related Work}
\label{sec:related}

In this section, we discuss previous work on detecting malicious accounts on social media, and survey research on disinformation carried out by troll accounts on social media.

\subsection{Detecting Malicious Activity on Social\\Media}

\descr{Detecting malicious messages.}
Computer security researchers attempted to curb the problem of malicious content on social networks by detecting malicious messages automatically (e.g., spam).
Yardi et al.~\cite{yardi2009detecting} developed a tool to detect Twitter spammers who abuse trending topics.
Thomas et al. presented {\sc Monarch}~\cite{thomas2011design}, a system that analyzes the URLs shared by social accounts for signs of maliciousness. 
Lee and Kim~\cite{lee2012warningbird} proposed {\sc WarningBird}, a system that analyzes correlated redirection chains of URLs in a number of URLs posted on Twitter to identify malicious tweets. 
Another line of work leverages clustering techniques to group together similar messages posted on social media and flagging them as spam~\cite{gao2010detecting, grier2010spam}. 
Liu et al.~\cite{liu2016detecting} calculated the topics shared by spammers with LDA, and then employed supervised learning to identify spammers based on the topics that they discussed. 

\descr{Detecting malicious accounts.}
Another approach is to identify malicious accounts that are active on social networks based on their characteristics.
Early work looked at characteristics that are typical of fake accounts, e.g., having an abnormal fraction of friends compared to followers, or posting content that was similar to each other~\cite{benevenuto2010detecting,stringhini2010detecting}. %
Yang et al. improved on this, identifying features in fake accounts that are more resilient to evasion by adversaries~\cite{yang2011free}.
Ghosh et al.~\cite{ghosh2012understanding} investigated link farming, a phenomenon used by spam accounts to allow them to receive a large number of followers.
Viswanath et al.~\cite{viswanath2014towards} applied Principal Components Analysis (PCA) to find patterns among features extracted from spam accounts. 
Egele et al.~\cite{egele2015towards} focused on detecting legitimate accounts that have been compromised by an adversary, showing that normal users have almost stable habits over time and that sudden anomalies in these habits are highly indicative of a compromise. 
Wang et al.~\cite{wang2013you} analyzed user click patterns to create user profiles and identify fake accounts using both supervised and unsupervised learning. 
\rev{Galan-Garcia et al.~\cite{galangarcia} aim at detecting fake accounts that harass social media users by analyzing the content of the comments made by those accounts.}

Other work is based on the assumption that fake accounts present fundamentally different social connections than real accounts.
Cai et al.~\cite{cai2012latent} split a social network into communities and tried to identify communities that connect in an unnatural or inconsistent way with the rest of the social network. 
Danezis et al.~\cite{danezis2009sybilinfer} used the same idea to detect compromised accounts, using a Bayesian Inference approach. 

Another line of research deals with the fact that fake accounts are commonly controlled by a single entity, and are therefore likely to act in a synchronized fashion.
Cao et al. proposed SynchroTrap~\cite{cao2014uncovering}, a detection system that clusters malicious accounts according to their actions and the time at which they are made. 
Stringhini et al. proposed {\sc EvilCohort}~\cite{stringhini2015evilcohort}, a system that identifies sets of social network accounts used by botnets, by looking at communities of accounts that are accessed by a common set of IP addresses.

\descr{Message Propagation.}
The third line of work focuses on the way in which messages propagate on social networks.
The assumption is that malicious messages (e.g., spam) will show different propagation patterns than legitimate ones.
Ye and Wu~\cite{ye2010measuring} studied propagation patterns of general messages and breaking news in Twitter, identifying patterns that are indicative of false or true information. 
Vosoughi et al. found that false news gets shared at a higher rate than true information~\cite{vosoughi2018spread}.
Weng et al.~\cite{weng2013virality} analyzed Twitter hashtags and showed that network communities can help to predict viral memes. 
Nematzadeh et al.~\cite{nematzadeh2014optimal} demonstrated that  
strong communities with high modularity can facilitate global diffusion by enhancing local, intra-community spreading. 
Xu et al.~\cite{xu2010toward} presented an early warning worm detection system that monitors the behavior of users to collect suspicious worm propagation evidence.
Through a simulation, Mezzour et al.~\cite{mezzour2014spam} showed how the diffusion of messages by hacked accounts differs from normal accounts. 
In particular, these accounts keep posting their content regardless of the engagement or feedback that they receive from other users.

\subsection{Bot and Troll Activity on Social Media}

A large body of work focused on social bots~\cite{bessi2016social,davis2016bot,ferrara2016rise,ferrara2017disinformation,varol2017online} and their role in spreading political disinformation, highlighting that bots can manipulate the public's opinion at a large scale, thus potentially affecting the outcome of elections.

Zhang et al.~\cite{zhang2021russianira} analyzed Russian Internet Research Agency (IRA)'s disinformation campaign on Twitter.
They emphasized that, in already polarized discussion topics such as politics, it is extremely challenging to distinguish between ``legitimate'' political expression and ``disinformation'' since such discussions are highly opinionated making them ideal targets for disinformation attacks.
Kumar et al.~\cite{kumar2017army} measured the phenomenon of multiple accounts controlled by the same user, called \emph{sockpuppets}, noting that these accounts actively attempt to manipulate users' opinions on online communities.

Mihaylov and Nakov~\cite{mihaylov2016hunting} identified two types of trolls: those who act on their own and those who are paid to spread specific messages.
In a related research effort, Mihaylov et al.~\cite{mihaylov2015finding} showed that trolls can indeed manipulate users' opinions in online forums.
Steward et al.~\cite{steward2018examining} studied the activity of Russian-sponsored trolls in the Black Lives Matter debate on Twitter.
They found that trolls infiltrated both left and right-leaning communities, with the goal of pushing specific narratives.
Varol et al.~\cite{varol2017early} developed a system to identify memes (ideas) that become popular due to coordinated efforts.
Ratkiewicz et al.~\cite{ratkiewicz2011detecting} used machine learning to detect the spread of false political information on Twitter.

Howard and Kollanyi~\cite{howard2016bots} found that the bots active during the 2016 Brexit referendum campaign were mostly pushing narratives that favored Brexit, with 1\% of the accounts generating 33\% of the overall messages.
Hegelich and Janetzko~\cite{hegelich2016are} investigated whether bots on Twitter are used as political actors.
By exposing and analyzing 1.7K bots on Twitter during the Russia-Ukraine conflict, they uncover their political agenda and show that bots exhibit various behaviors, e.g., trying to hide their identity, promoting topics through the use of hashtags, and retweeting messages with particularly interesting content.
Badawy et al.~\cite{badawy2018falls} aim to predict users that are likely to spread information from state-sponsored actors, while Dutt et al.~\cite{dutt2018senator} focus on the Facebook platform and analyze ads shared by Russian trolls to find the cues that make them effective.

Zannettou et al. analyzed state-sponsored troll accounts active on Twitter and Reddit between 2014 and 2018~\cite{zannettou2019disinformation,zannettou2019let}.
They found that these accounts were created in waves, and measured their efficiency in spreading their content on those platforms as well as on other Web communities.
In follow-up work, the same authors presented an analysis pipeline to study the images posted by these accounts on Twitter~\cite{zannettou2019characterizing}.

The work discussed above predominantly focuses on studying the activity of, rather than detecting, troll accounts.
Closer to our work are the very few efforts toward detection~\cite{volkova2016account,luceri2020detecting}.
Volkova and Bell~\cite{volkova2016account} analyzed 180k Twitter accounts that were active during the Russia-Ukraine conflict, finding that lexical features are highly predictive of whether an account will be identified as a troll by Twitter and suspended. 
\rev{
Luceri et al.~\cite{luceri2020detecting} apply Inverse Reinforcement Learning (IRL) to detect trolls.
They use similar features to those previously used by bot detection systems (e.g., replies, retweets) to automatically detect troll accounts on Twitter.
They find that troll accounts on Twitter keep posting regardless of whether other users react to these posts, while the activity of regular users is influenced by these interactions. 
While a detector trained this way might work on a carefully selected dataset, the behavior that Luceri et al. model is not specific to troll accounts, but it rather matches any account operating in an automated fashion.
This means that not only spam accounts would be flagged, but also auto-moderator bots that are commonly deployed on Reddit. 
Therefore, we argue that this approach is not suitable for the problem at hand.} 
Unlike their work, our effort is the first to look at coordination between troll accounts and leverage interaction between them for detection.
\newrev{Weller et al.~\cite{identifying_trolls} use deep learning (i.e., Convolutional Neural Networks and Region-based Convolutional Neural Networks) to detect Russian trolls based on their comments.
This is to the best of our knowledge, the only research work that uses the same dataset as ours for detection.
Their performance of their validation experiments on the ground truth dataset is substantially lower than \system, which further highlights the utility of our approach.
}

\section{Conclusion}
This paper presented \system,  a system that learns the typical behavior of known state-sponsored troll accounts on Reddit with the goal of finding more such accounts.
The core insight behind \system is that troll accounts tend to interact with each other to further disinformation narratives and to polarize online discussion.
We tested \system on a Reddit dataset\edc{which dataset} and identified \detectedtrolls potential troll accounts.
\rev{
We find that 66\% of the detected accounts show signs of being controlled by malicious actors, and that these accounts as a group show signs of synchronization with the set of known troll accounts, including using similar language.
}
Overall, we are confident that our findings can serve as a promising starting point for researchers and online social networks to develop more effective detection systems against disinformation actors.

\descr{Acknowledgments.} We thank the anonymous reviewers for their comments and the discussion during the interactive rebuttal phase. 
This paper was supported by the NSF under grants CNS-1942610, IIS-2046590, CNS-2114407, IIP-1827700, and CNS-2114411, as well as the UK's National Research Centre on Privacy, Harm Reduction, and Adversarial Influence Online (REPHRAIN, UKRI grant: EP/V011189/1).

\small
\bibliographystyle{abbrv}

\begin{thebibliography}{10}

\bibitem{badawy2018falls}
A.~Badawy, K.~Lerman, and E.~Ferrara.
\newblock {Who Falls for Online Political Manipulation?}
\newblock {\em arXiv:1808.03281}, 2018.

\bibitem{baumgartner2020pushshift}
J.~Baumgartner, S.~Zannettou, B.~Keegan, M.~Squire, and J.~Blackburn.
\newblock {The Pushshift Reddit Dataset}.
\newblock In {\em AAAI International Conference on Web and Social Media
  (ICWSM)}, 2020.

\bibitem{benevenuto2010detecting}
F.~Benevenuto, G.~Magno, T.~Rodrigues, and V.~Almeida.
\newblock Detecting spammers on twitter.
\newblock In {\em Collaboration, electronic messaging, anti-abuse and spam
  conference (CEAS)}, 2010.

\bibitem{bessi2016social}
A.~Bessi and E.~Ferrara.
\newblock {Social bots distort the 2016 US Presidential election online
  discussion}.
\newblock {\em First Monday}, 21(11), 2016.

\bibitem{Blondel2008}
V.~D. Blondel, J.-L. Guillaume, R.~Lambiotte, and E.~Lefebvre.
\newblock Fast unfolding of communities in large networks.
\newblock {\em Journal of Statistical Mechanics: Theory and Experiment},
  2008(10):P10008, Oct 2008.

\bibitem{breiman2001random}
L.~Breiman.
\newblock {Random forests}.
\newblock {\em Machine Learning}, 45(1), 2001.

\bibitem{cai2012latent}
Z.~Cai and C.~Jermaine.
\newblock The latent community model for detecting sybil attacks in social
  networks.
\newblock In {\em ISOC Network and Distributed Systems Security Symposium
  (NDSS)}, 2012.

\bibitem{cao2014uncovering}
Q.~Cao, X.~Yang, J.~Yu, and C.~Palow.
\newblock Uncovering large groups of active malicious accounts in online social
  networks.
\newblock 2014.

\bibitem{danezis2009sybilinfer}
G.~Danezis and P.~Mittal.
\newblock Sybilinfer: Detecting sybil nodes using social networks.
\newblock In {\em ISOC Network and Distributed Systems Security Symposium
  (NDSS)}, 2009.

\bibitem{davis2016bot}
C.~A. Davis, O.~Varol, E.~Ferrara, A.~Flammini, and F.~Menczer.
\newblock {BotOrNot: A System to Evaluate Social Bots}.
\newblock In {\em The Web Conference (WWW)}, 2016.

\bibitem{deng2016efficient}
Z.~Deng, X.~Zhu, D.~Cheng, M.~Zong, and S.~Zhang.
\newblock {Efficient kNN classification algorithm for big data}.
\newblock {\em Neurocomputing}, 195, 2016.

\bibitem{dutt2018senator}
R.~Dutt, A.~Deb, and E.~Ferrara.
\newblock {'Senator, We Sell Ads': Analysis of the 2016 Russian Facebook Ads
  Campaign}.
\newblock {\em arXiv:1809.10158}, 2018.

\bibitem{egele2015towards}
M.~Egele, G.~Stringhini, C.~Kruegel, and G.~Vigna.
\newblock Towards detecting compromised accounts on social networks.
\newblock {\em {Transactions on Dependable and Secure Computing (TDSC)}}, 2015.

\bibitem{ferrara2017disinformation}
E.~Ferrara.
\newblock {Disinformation and social bot operations in the run up to the 2017
  French presidential election}.
\newblock {\em ArXiv 1707.00086}, 2017.

\bibitem{ferrara2016rise}
E.~Ferrara, O.~Varol, C.~Davis, F.~Menczer, and A.~Flammini.
\newblock The rise of social bots.
\newblock {\em Communications of the ACM}, 2016.

\bibitem{galangarcia}
P.~Gal{\'a}n-Garc{\'\i}a, J.~G. d.~l. Puerta, C.~L. G{\'o}mez, I.~Santos, and
  P.~G. Bringas.
\newblock Supervised machine learning for the detection of troll profiles in
  twitter social network: Application to a real case of cyberbullying.
\newblock {\em Logic Journal of the IGPL}, 24(1):42--53, 2016.

\bibitem{gao2010detecting}
H.~Gao, J.~Hu, C.~Wilson, Z.~Li, Y.~Chen, and B.~Y. Zhao.
\newblock Detecting and characterizing social spam campaigns.
\newblock In {\em ACM Internet Measurement Conference (IMC)}, 2010.

\bibitem{gensim}
Gensim.
\newblock {Word2vec embeddings}.
\newblock \url{https://radimrehurek.com/gensim/models/word2vec.html}.

\bibitem{ghosh2012understanding}
S.~Ghosh, B.~Viswanath, F.~Kooti, N.~K. Sharma, G.~Korlam, F.~Benevenuto,
  N.~Ganguly, and K.~P. Gummadi.
\newblock Understanding and combating link farming in the twitter social
  network.
\newblock In {\em The Web Conference (WWW)}, 2012.

\bibitem{grier2010spam}
C.~Grier, K.~Thomas, V.~Paxson, and M.~Zhang.
\newblock {@spam: The Underground on 140 Characters or Less}.
\newblock In {\em ACM Conference on Computer and Communications Security
  (CCS)}, 2010.

\bibitem{hearst1998support}
M.~A. {Hearst}, S.~T. {Dumais}, E.~{Osuna}, J.~{Platt}, and B.~{Scholkopf}.
\newblock Support vector machines.
\newblock {\em IEEE Intelligent Systems and their Applications}, 13(4):18--28,
  1998.

\bibitem{hegelich2016are}
S.~Hegelich and D.~Janetzko.
\newblock {Are Social Bots on Twitter Political Actors? Empirical Evidence from
  a Ukrainian Social Botnet}.
\newblock In {\em AAAI International Conference on Web and Social Media
  (ICWSM)}, 2016.

\bibitem{identifying_trolls}
J.~W. Henry~Weller.
\newblock {Identifying Russian Trolls on Reddit with Deep Learning and BERT
  Word Embeddings}.
\newblock \url{https://web.stanford.edu/class/cs224n/posters/15739845.pdf},
  2019.

\bibitem{howard2016bots}
P.~N. Howard and B.~Kollanyi.
\newblock {Bots, \#StrongerIn, and \#Brexit: Computational Propaganda during
  the UK-EU Referendum}.
\newblock {\em CoRR}, abs/1606.06356, 2016.

\bibitem{gephi}
M.~Jacomy, T.~Venturini, S.~Heymann, and M.~Bastian.
\newblock Forceatlas2, a continuous graph layout algorithm for handy network
  visualization designed for the gephi software.
\newblock {\em PLOS ONE}, 9(6):1--12, 06 2014.

\bibitem{kulkarni2014statistically}
V.~Kulkarni, R.~Al-Rfou, B.~Perozzi, and S.~Skiena.
\newblock Statistically significant detection of linguistic change, 2014.

\bibitem{kumar2017army}
S.~Kumar, J.~Cheng, J.~Leskovec, and V.~Subrahmanian.
\newblock An army of me: Sockpuppets in online discussion communities.
\newblock In {\em The Web Conference (WWW)}, 2017.

\bibitem{kwak2010twitter}
H.~Kwak, C.~Lee, H.~Park, and S.~Moon.
\newblock What is twitter, a social network or a news media?
\newblock In {\em The Web Conference (WWW)}, 2010.

\bibitem{tfidf}
M.~J. Lavin.
\newblock {Analyzing Documents with TF-IDF}.
\newblock
  \url{https://programminghistorian.org/en/lessons/analyzing-documents-with-tfidf},
  2019.

\bibitem{lee2012warningbird}
S.~Lee and J.~Kim.
\newblock Warningbird: Detecting suspicious urls in twitter stream.
\newblock In {\em ISOC Network and Distributed Systems Security Symposium
  (NDSS)}, 2012.

\bibitem{lerman2010information}
K.~Lerman and R.~Ghosh.
\newblock Information contagion: An empirical study of the spread of news on
  digg and twitter social networks.
\newblock In {\em AAAI International Conference on Web and Social Media
  (ICWSM)}, 2010.

\bibitem{liu2016detecting}
L.~Liu, Y.~Lu, Y.~Luo, R.~Zhang, L.~Itti, and J.~Lu.
\newblock Detecting" smart" spammers on social network: A topic model approach.
\newblock {\em arXiv:1604.08504}, 2016.

\bibitem{luceri2020detecting}
L.~Luceri, S.~Giordano, and E.~Ferrara.
\newblock Detecting troll behavior via inverse reinforcement learning: A case
  study of russian trolls in the 2016 us election.
\newblock In {\em AAAI International Conference on Web and Social Media
  (ICWSM)}, 2020.

\bibitem{mezzour2014spam}
G.~Mezzour and K.~M. Carley.
\newblock Spam diffusion in a social network initiated by hacked e--mail
  accounts.
\newblock {\em International Journal of Security and Networks}, 9(3):144--153,
  2014.

\bibitem{mihaylov2015finding}
T.~Mihaylov, G.~Georgiev, and P.~Nakov.
\newblock {Finding Opinion Manipulation Trolls in News Community Forums}.
\newblock In {\em CoNLL}, 2015.

\bibitem{mihaylov2016hunting}
T.~Mihaylov and P.~Nakov.
\newblock {Hunting for Troll Comments in News Community Forums}.
\newblock In {\em ACL}, 2016.

\bibitem{mueller2019mueller}
R.~S. Mueller.
\newblock The mueller report: Report on the investigation into russian
  interference in the 2016 presidential election.
\newblock \url{https://www.justice.gov/archives/sco/file/1373816/download},
  2019.

\bibitem{nic2021foreign}
{National Intelligence Council}.
\newblock Foreign threats to the 2020 us federal elections.
\newblock
  \url{https://www.dni.gov/files/ODNI/documents/assessments/ICA-declass-16MAR21.pdf},
  2021.

\bibitem{nematzadeh2014optimal}
A.~Nematzadeh, E.~Ferrara, A.~Flammini, and Y.-Y. Ahn.
\newblock Optimal network modularity for information diffusion.
\newblock {\em Physical review letters}, 2014.

\bibitem{nilizadeh2017poised}
S.~Nilizadeh, F.~Labr{\`e}che, A.~Sedighian, A.~Zand, J.~Fernandez, C.~Kruegel,
  G.~Stringhini, and G.~Vigna.
\newblock {POISED: Spotting Twitter Spam Off the Beaten Paths}.
\newblock In {\em ACM Conference on Computer and Communications Security
  (CCS)}, 2017.

\bibitem{praw}
PRAW.
\newblock {The Python Reddit API Wrapper}.
\newblock \url{https://praw.readthedocs.io/en/latest/}.

\bibitem{ratkiewicz2011detecting}
J.~Ratkiewicz, M.~Conover, M.~R. Meiss, B.~Gonçalves, A.~Flammini, and
  F.~Menczer.
\newblock {Detecting and Tracking Political Abuse in Social Media}.
\newblock In {\em AAAI International Conference on Web and Social Media
  (ICWSM)}, 2011.

\bibitem{reddittrolls}
Reddit.
\newblock {Reddit's 2017 transparency report and suspect account findings}.
\newblock
  \url{https://www.reddit.com/r/announcements/comments/8bb85p/reddits\_2017\_transparency\_report\_and\_suspect/},
  2017.

\bibitem{samory2018conspiracies}
M.~Samory and T.~Mitra.
\newblock Conspiracies online: User discussions in a conspiracy community
  following dramatic events.
\newblock In {\em AAAI International Conference on Web and Social Media
  (ICWSM)}, 2018.

\bibitem{starbird2017examining}
K.~Starbird.
\newblock Examining the alternative media ecosystem through the production of
  alternative narratives of mass shooting events on twitter.
\newblock In {\em AAAI International Conference on Web and Social Media
  (ICWSM)}, 2017.

\bibitem{starbird2019disinformation}
K.~Starbird, A.~Arif, and T.~Wilson.
\newblock Disinformation as collaborative work: Surfacing the participatory
  nature of strategic information operations.
\newblock {\em Proceedings of the ACM on Human-Computer Interaction}, 2019.

\bibitem{steward2018examining}
L.~Steward, A.~Arif, and K.~Starbird.
\newblock {Examining Trolls and Polarization with a Retweet Network}.
\newblock In {\em MIS2}, 2018.

\bibitem{stringhini2010detecting}
G.~Stringhini, C.~Kruegel, and G.~Vigna.
\newblock Detecting spammers on social networks.
\newblock In {\em Annual Computer Security Applications Conference (ACSAC)},
  2010.

\bibitem{stringhini2015evilcohort}
G.~Stringhini, P.~Mourlanne, G.~Jacob, M.~Egele, C.~Kruegel, and G.~Vigna.
\newblock $\{$EVILCOHORT$\}$: Detecting communities of malicious accounts on
  online services.
\newblock In {\em USENIX Security Symposium}, 2015.

\bibitem{swain1977decision}
P.~H. Swain and H.~Hauska.
\newblock {The decision tree classifier: Design and potential}.
\newblock {\em IEEE Transactions on Geoscience Electronics}, 15(3), 1977.

\bibitem{thomas2011design}
K.~Thomas, C.~Grier, J.~Ma, V.~Paxson, and D.~Song.
\newblock Design and evaluation of a real-time url spam filtering service.
\newblock In {\em IEEE Symposium on Security and Privacy}, 2011.

\bibitem{twittertrolls}
Twitter.
\newblock {Information Operations}.
\newblock
  \url{https://transparency.twitter.com/en/reports/information-operations.html},
  2019.

\bibitem{varol2017online}
O.~Varol, E.~Ferrara, C.~A. Davis, F.~Menczer, and A.~Flammini.
\newblock {Online Human-Bot Interactions: Detection, Estimation, and
  Characterization}.
\newblock In {\em AAAI International Conference on Web and Social Media
  (ICWSM)}, 2017.

\bibitem{varol2017early}
O.~Varol, E.~Ferrara, F.~Menczer, and A.~Flammini.
\newblock {Early detection of promoted campaigns on social media}.
\newblock {\em EPJ Data Science}, 2017.

\bibitem{viswanath2014towards}
B.~Viswanath, M.~A. Bashir, M.~Crovella, S.~Guha, K.~P. Gummadi,
  B.~Krishnamurthy, and A.~Mislove.
\newblock Towards detecting anomalous user behavior in online social networks.
\newblock In {\em USENIX Security Symposium}, 2014.

\bibitem{rfe}
D.~Volcheck.
\newblock {One Professional Russian Troll Tells All}.
\newblock
  \url{https://www.rferl.org/a/how-to-guide-russian-trolling-trolls/26919999.html},
  2015.

\bibitem{volkova2016account}
S.~Volkova and E.~Bell.
\newblock {Account Deletion Prediction on RuNet: A Case Study of Suspicious
  Twitter Accounts Active During the Russian-Ukrainian Crisis}.
\newblock In {\em NAACL-HLT}, 2016.

\bibitem{vosoughi2018spread}
S.~Vosoughi, D.~Roy, and S.~Aral.
\newblock The spread of true and false news online.
\newblock {\em Science}, 359(6380):1146--1151, 2018.

\bibitem{wang2013you}
G.~Wang, T.~Konolige, C.~Wilson, X.~Wang, H.~Zheng, and B.~Y. Zhao.
\newblock You are how you click: Clickstream analysis for sybil detection.
\newblock In {\em USENIX Security Symposium}, 2013.

\bibitem{wang2020understanding}
Y.~Wang, F.~Tamahsbi, J.~Blackburn, B.~Bradlyn, E.~De~Cristofaro, D.~Magerman,
  S.~Zannettou, and G.~Stringhini.
\newblock Understanding the use of fauxtography on social media.
\newblock In {\em AAAI International Conference on Web and Social Media
  (ICWSM)}, 2021.

\bibitem{weng2013virality}
L.~Weng, F.~Menczer, and Y.-Y. Ahn.
\newblock Virality prediction and community structure in social networks.
\newblock {\em Scientific reports}, 3, 2013.

\bibitem{weninger2014exploration}
T.~Weninger.
\newblock An exploration of submissions and discussions in social news: Mining
  collective intelligence of reddit.
\newblock {\em Social Network Analysis and Mining}, 2014.

\bibitem{weninger2013exploration}
T.~Weninger, X.~A. Zhu, and J.~Han.
\newblock An exploration of discussion threads in social news sites: A case
  study of the reddit community.
\newblock In {\em {ASONAM}}, 2013.

\bibitem{xiadisinfo}
Y.~Xia, J.~Lukito, Y.~Zhang, C.~Wells, S.~J. Kim, and C.~Tong.
\newblock Disinformation, performed: Self-presentation of a russian ira account
  on twitter.
\newblock {\em Information, Communication and Society}, 2019.

\bibitem{xudeep}
T.~Xu, G.~Goossen, H.~K. Cevahir, S.~Khodeir, Y.~Jin, F.~Li, S.~Shan, S.~Patel,
  D.~Freeman, and P.~Pearce.
\newblock Deep entity classification: Abusive account detection for online
  social networks.
\newblock In {\em USENIX Security Symposium}, 2021.

\bibitem{xu2010toward}
W.~Xu, F.~Zhang, and S.~Zhu.
\newblock Toward worm detection in online social networks.
\newblock In {\em Annual Computer Security Applications Conference (ACSAC)},
  2010.

\bibitem{yang2011free}
C.~Yang, R.~C. Harkreader, and G.~Gu.
\newblock Die free or live hard? empirical evaluation and new design for
  fighting evolving twitter spammers.
\newblock In {\em International Workshop on Recent Advances in Intrusion
  Detection}, 2011.

\bibitem{yardi2009detecting}
S.~Yardi, D.~Romero, G.~Schoenebeck, and d.~boyd.
\newblock Detecting spam in a twitter network.
\newblock {\em First Monday}, 2009.

\bibitem{ye2010measuring}
S.~Ye and S.~F. Wu.
\newblock {\em Measuring message propagation and social influence on Twitter}.
\newblock Springer, 2010.

\bibitem{yuan2019detecting}
D.~Yuan, Y.~Miao, N.~Z. Gong, Z.~Yang, Q.~Li, D.~Song, Q.~Wang, and X.~Liang.
\newblock Detecting fake accounts in online social networks at the time of
  registrations.
\newblock In {\em ACM Conference on Computer and Communications Security
  (CCS)}, 2019.

\bibitem{zannettou2019characterizing}
S.~Zannettou, B.~Bradlyn, E.~De~Cristofaro, G.~Stringhini, and J.~Blackburn.
\newblock Characterizing the use of images by state-sponsored troll accounts on
  twitter.
\newblock In {\em AAAI International Conference on Web and Social Media
  (ICWSM)}, 2019.

\bibitem{zannettou2017web}
S.~Zannettou, T.~Caulfield, E.~De~Cristofaro, N.~Kourtellis, I.~Leontiadis,
  M.~Sirivianos, G.~Stringhini, and J.~Blackburn.
\newblock {The Web Centipede: Understanding How Web Communities Influence Each
  Other Through the Lens of Mainstream and Alternative News Sources}.
\newblock In {\em ACM Internet Measurement Conference (IMC)}, 2017.

\bibitem{zannettou2019disinformation}
S.~Zannettou, T.~Caulfield, E.~De~Cristofaro, M.~Sirivianos, G.~Stringhini, and
  J.~Blackburn.
\newblock Disinformation warfare: Understanding state-sponsored trolls on
  twitter and their influence on the web.
\newblock In {\em {WWW Companion}}, 2019.

\bibitem{zannettou2019let}
S.~Zannettou, T.~Caulfield, W.~Setzer, M.~Sirivianos, G.~Stringhini, and
  J.~Blackburn.
\newblock Who let the trolls out?: Towards understanding state-sponsored
  trolls.
\newblock In {\em {ACM Conference on Web Science}}, 2019.

\bibitem{zannettou2019quantitative}
S.~Zannettou, J.~Finkelstein, B.~Bradlyn, and J.~Blackburn.
\newblock {A Quantitative Approach to Understanding Online Antisemitism}.
\newblock In {\em AAAI International Conference on Web and Social Media
  (ICWSM)}, 2020.

\bibitem{zhang2021russianira}
Y.~Zhang, J.~Lukito, M.-H. Su, J.~Suk, Y.~Xia, S.~J. Kim, L.~Doroshenko, and
  C.~Wells.
\newblock Assembling the networks and audiences of disinformation: How
  successful russian ira twitter accounts built their followings, 2015–2017.
\newblock {\em Journal of Communication}, 01 2021.

\end{thebibliography}
\end{document}